\makeatletter
\def\cl@chapter{}
\makeatother

\documentclass[smallextended]{svjour3}       
\smartqed  
\usepackage{scalerel}
\usepackage{tikz}
\usetikzlibrary{svg.path}

\definecolor{orcidlogocol}{HTML}{A6CE39}
\tikzset{
  orcidlogo/.pic={
    \fill[orcidlogocol] svg{M256,128c0,70.7-57.3,128-128,128C57.3,256,0,198.7,0,128C0,57.3,57.3,0,128,0C198.7,0,256,57.3,256,128z};
    \fill[white] svg{M86.3,186.2H70.9V79.1h15.4v48.4V186.2z}
                 svg{M108.9,79.1h41.6c39.6,0,57,28.3,57,53.6c0,27.5-21.5,53.6-56.8,53.6h-41.8V79.1z M124.3,172.4h24.5c34.9,0,42.9-26.5,42.9-39.7c0-21.5-13.7-39.7-43.7-39.7h-23.7V172.4z}
                 svg{M88.7,56.8c0,5.5-4.5,10.1-10.1,10.1c-5.6,0-10.1-4.6-10.1-10.1c0-5.6,4.5-10.1,10.1-10.1C84.2,46.7,88.7,51.3,88.7,56.8z};
  }
}

\newcommand\orcidicon[1]{\href{https://orcid.org/#1}{\mbox{\scalerel*{
\begin{tikzpicture}[yscale=-1,transform shape]
\pic{orcidlogo};
\end{tikzpicture}
}{|}}}}
\usepackage{tipa}
\usepackage{natbib}
\setcitestyle{aysep={}} 
\usepackage[colorlinks=true,linkcolor=black,anchorcolor=black,citecolor=black,filecolor=black,menucolor=black,runcolor=black,urlcolor=black]{hyperref}
\usepackage{cleveref}
\usepackage{adjustbox}
\usepackage{booktabs}
\usepackage{subfig}
\usepackage{pifont}
\usepackage[framemethod=TikZ]{mdframed}
\usepackage{enumitem}
\usepackage[misc]{ifsym}

\def\BibTeX{{\rm B\kern-.05em{\sc i\kern-.025em b}\kern-.08emT\kern-.1667em\lower.7ex\hbox{E}\kern-.125emX}}

\newcommand{\ahref}[2]{\href{#1}{\nolinkurl{#2}}}

\newtheorem{inner}{\innerheader}
\newcommand{\innerheader}{} \newenvironment{defi}[1]
{\renewcommand\innerheader{#1}\begin{inner}} {\end{inner}}

\newcommand*\rectangled[1]{\tikz[baseline=(char.base)]{\node[shape=rectangle,draw,inner sep=2pt] (char) {#1};}}

\usepackage{xspace}

\newcommand{\ename}{\textsc{Pr\"azi}\xspace}

\newcommand{\npm}{\textsc{npm}\xspace}
\newcommand\maven{\textsc{Maven}\xspace}
\newcommand\nuget{\textsc{NuGet}\xspace}

\newcommand\cpan{\textsc{CPAN}\xspace}
\newcommand\cran{\textsc{CRAN}\xspace}

\newcommand\crates{\textsc{Crates.io}\xspace}
\newcommand\mavencentral{\textsc{Maven Central}\xspace}
\newcommand\cargo{\textsc{Cargo}\xspace}
\newcommand\cargotoml{\textsc{Cargo.toml}\xspace}

\newcommand\github{\textsc{GitHub}\xspace}

\journalname{Empirical Software Engineering}

\begin{document}

\title{\ename: From Package-based to Call-based Dependency Networks
}

\author{Joseph Hejderup \protect\orcidicon{0000-0001-8626-9964} \and Moritz Beller*\thanks{*Work largely conducted while the author was a researcher at TU Delft, The Netherlands.} \protect\orcidicon{0000-0003-4852-0526} \and Konstantinos Triantafyllou \and Georgios Gousios* \protect\orcidicon{0000-0002-8495-7939}}

\authorrunning{Joseph Hejderup et al.}

\institute{\Letter~Joseph Hejderup  \at
           Delft University of Technology \\
           \email{j.i.hejderup@tudelft.nl}           
           \and
           Moritz Beller \at
           Facebook, Inc. \\
           \email{mmb@fb.com} 
           \and
           Konstantinos Triantafyllou \at
           University of Athens \\
           \email{ks.triantafyllou@gmail.com}
           \and
           Georgios Gousios  \at
           Facebook, Inc. \\
           \email{gousiosg@fb.com} 
}

\date{Received: date / Accepted: date}

\maketitle

\begin{abstract}
\begin{sloppypar}
Modern programming languages such as Java, JavaScript, and Rust encourage
software reuse by hosting diverse and fast-growing repositories of highly
interdependent packages (i.e., reusable libraries) for their users. The standard way to study
the interdependence between software packages is to infer a package dependency
network by parsing manifest data. Such networks help answer questions such as
``How many packages have dependencies to packages with known security issues?''
or ``What are the most used packages?''. However, an overlooked aspect in
existing studies is that manifest-inferred relationships do not necessarily
examine the actual usage of these dependencies in source code. To better model
dependencies between packages, we developed \ename, an approach combining
manifests and call graphs of packages. \ename constructs a dependency network at
the more fine-grained function-level, instead of at the manifest level. This
paper discusses a prototypical \ename implementation for the popular system
programming language Rust. We use \ename to characterize Rust's package
repository, \crates, at the function level and perform a comparative study with
metadata-based networks. Our results show that metadata-based networks
generalize how packages use their dependencies. Using \ename, we find
packages call only 40\% of their resolved dependencies, and that manual analysis
of 34 cases reveals that not all packages use a dependency the same way. We
argue that researchers and practitioners interested in understanding how
developers or programs use dependencies should account for its context---not the
sum of all resolved dependencies.

\end{sloppypar}
\end{abstract}

\section{Introduction}
\begin{sloppypar}
Converting information between different well-known formats, accessing external
storage, manipulating information such as numbers, locations, and dates, or
integrating with popular online services are examples of essential operations
that developers need to handle in software projects. Unlike the standard library
of programming languages, these essential operations change over time
as a result of evolving technologies (e.g., shift from XML to JSON) or provide
support to niche user communities (e.g., interfaces to Twitter API or Amazon AWS
SDK).  In addition to a standard library, modern programming languages such as
Java, JavaScript, C\#, and Rust also host public repositories for developers to contribute essential operations in the form of reusable
libraries (also known as packages). A package manager such as \maven, \npm,
\nuget, and \cargo enables developers to discover and import packages from
repositories in their workspace.

To be modular, a package should perform a
well-defined task, developed with simple interfaces, and be pluggable
(composable) with other packages~\citep{npm:unix:design,abdalkareem2019impact}.
The manifest file such as Rust's \cargotoml and \npm's \texttt{package.json} in
every package makes libraries composable: developers declare in the manifest how
others can import their library and also if it utilizes external libraries by
specifying dependencies on other existing packages. As packages can depend on
one another through manifests, package repositories implicitly form a complex
network, known as a Package Dependency
Network (PDN)~\citep{decan2018impact,hejderup2015dependencies,kikas2017structure}.

In light of repository-wide incidents such as the \texttt{left-pad} package
removal~\citep{npm:leftpad:online}, the hiding of a bitcoin wallet stealer in the
legitimate \texttt{event-stream} package~\citep{npm:eventstream}, and malicious
typosquatting packages in PyPI~\citep{pypi:typo}, researchers are conducting
network analysis of package repositories for risk
assessment~\citep{zimmermann2019small,decan2018impact,kikas2017structure},
sustainability evaluation~\citep{valiev2018ecosystem,decan2019empirical}, license
violations~\citep{duan2017identifying}, and for detecting breaking
changes~\citep{mezzetti2018type,chen2020taming,Mujahid_MSR2020}. Constructing a
PDN for such analyses typically involves mining
available manifests in the repository and then resolving dependency constraints
in each manifest using a specific resolver (i.e., variations of \texttt{semver})
to infer relationships between
packages~\citep{kikas2017structure,hejderup2015dependencies}.

Inferring networks solely from package manifests yields an incomplete
representation of package repositories. Manifests only describe metadata of
package dependencies and thus miss information on actual source code reuse,
making network analysis prone to false positives. For example, a project might have redundant dependencies
to packages whose functionality is not used anymore. Without knowing how packages actually use external
libraries,~\cite{ponta2018beyond} and~\cite{zapata2018towards}'s work on vulnerability checking packages
demonstrates that metadata-based analysis have limited actionability, making it
difficult for developers to understand how vulnerabilities in external libraries
affect their code. Increasingly, package repository workgroups such as the Rust
Ecosystem WG\footnote{\ahref{http://web.archive.org/web/20201201224020/https://github.com/rust-lang-nursery/ecosystem-wg}{https://github.com/rust-lang-nursery/ecosystem-wg}}
are also calling for more comprehensive network analysis of package repositories
to support code-centric analysis for more effective identification of critical
yet unstable packages~\citep{zhang2020enabling,bogart2016break}. One such
example is the \texttt{Libz Blitz}~\citep{rust:lib:blitz} initiative where
community members come together and contribute to poorly maintained yet critical
packages in \crates as an effort to stabilize highly reused code in the
repository.

This work proposes code-centric dependency network analysis by inferring
dependency relationships at the function call level. Call graphs 
capture how functions between packages use each other and thus naturally lend
themselves to this objective. We coin networks generated from call graphs
Call-based Dependency Networks (CDNs). To generate a CDN from a package
repository, we devise \ename, an approach that generates call graphs
of packages and then merges them into a single network with functions embedding
package qualifiers. The result is a more fine-grained dependency network that
improves over current PDN analyses by examining the actual package
dependencies in use. 

We implement \ename for \crates to demonstrate the feasibility of our approach. Unlike repositories hosting analyzable
binaries such as \mavencentral, \crates requires large-scale compilation of the
repository to produce binaries for call graph generation. The resulting CDN
comprises 90\% of all compilable packages, achieving a near-complete
representation of \crates. Then, inspired by~\cite{kikas2017structure}'s PDN
study, we characterize and derive new insights on the evolution of \crates. We also compare CDNs
against dependency networks derived from conventional
metadata to understand their differences and similarities for dependency network
analyses. Lastly, we manually investigate 34 direct and transitive package
relationships to understand how reliably a CDN represents actual use of
dependencies in the source code.

Our results find that one in two function calls in \crates are a call from a
package to a dependency, suggesting high code reuse. On average, we find that a package calls at least one
function in 78.8\% of its direct dependencies and at least one function in 40\% of its transitive
dependencies, suggesting that more than half of all transitive dependencies of packages are
potentially not called. When looking at APIs, packages have
three times more indirect (i.e., transitive) callers than direct callers. On average, a package has two new function calls every six months. Moreover, the number of calls from a package to its dependencies increases by 6.6 new direct calls and 12.2 indirect calls every six months. Reachability analysis
reveals that a majority of packages in \crates have no or limited
reachability. Only a handful packages (i.e., 0.37\% of packages in 2020) are
reachable by more than 10\% of \crates. Among the most central packages, the
most reachable
function can reach up to 30\% of all packages in \crates. The high indirect
use of APIs in transitive dependencies of packages could constitute an important
but missing confounding variable in API studies and manifest as an important threat to security and stability in practice.

The metadata-based networks and call-based networks report similar results for
analysis involving direct package relationships. However, notable differences exist between the studied networks for
transitive dependencies and for analyzing the most dependent packages. Metadata
networks report twice the number of transitive dependencies than the CDN. Our
findings in the manual analysis indicate that the high variance is a result of
transitive dependencies not being indirectly reachable (utilized) from the
package. A
package uses a subset of its direct dependencies---not all available
functionality. Thus, analysis of transitive dependencies is not generalizable
but contextual. Two packages that depend on the same library and
have two different use cases are likely to use their transitive dependencies
differently. Thus, dependency checkers, such as
\github's~\texttt{Dependabot}\footnote{\ahref{http://web.archive.org/web/20201201220347/https://dependabot.com/}{https://dependabot.com/}}
and
Rust's~\texttt{cargo-audit}\footnote{\ahref{http://web.archive.org/web/20201201224403/https://github.com/RustSec/cargo-audit}{https://github.com/RustSec/cargo-audit}},
should consider augmenting their recommendations with call graph information to
help developers make more informed decisions and reduce false
positives. As a step towards inferring networks from the source code of package
repositories, \ename can enable both
researchers and practitioners to estimate complex patterns of relationships
between packages and their functions. 

In summary, this work makes the following contributions: 

\begin{itemize}
   \item An approach to create call-based package dependency networks (CDNs) called \ename.
   \item An open-source implementation for generating CDN of Rust's \crates
   \item An empirical study describing the structure, evolution, and fragility of
   \crates from a package and function view.
   \item A comparison of network analyses using \ename CDN, metadata network, and compile-validated network.
   \item Two datasets for replication: CDNs for \crates and dataset of all generated call graphs.
\end{itemize}

For the reproducibility of our approach, generated CDNs, and study, we have made
the source code, the processing scripts and our data publicly available in a
replication package available~\citep{joseph_hejderup_2021_4478981}.
\end{sloppypar}

\section{Background}\label{sec:bg}
\subsection{Related Work}\label{sec:bg:rel}
Analyzing package repositories from a network perspective has become an
important research area in light of numerous incidents such as the removal of
the \texttt{left-pad} package in \npm and recent moves to emulate such problems
on package dependency
networks~\citep{kikas2017structure,kula2017modeling,decan2019empirical,zerouali2018empirical}.
The aftermath of the \texttt{left-pad} incident~\citep{npm:leftpad:online} in
2016 raised questions on how the removal of a single 11 LOC package downloaded
over $575,000$ times could break the build for large groups of seemingly
unrelated packages in \npm. To understand how certain packages exhibit such a
large degree of influence in package repositories,~\cite{kikas2017structure}'s
network analysis of three package repositories---\npm, \crates, and
RubyGems---uncovered that package repositories have scale-free network
properties~\citep{albert2002statistical}. As a result of a large number of
end-user applications depending on a popular set of packages (such as the
\texttt{babel} compiler), these popular yet distinct packages become hubs in
package dependency networks. Packages that act as hubs are not isolated
packages; they also depend on small and common utility packages such as
\texttt{left-pad} that appear as transitive dependencies for end-users. By
reversing the direction of package dependency
networks,~\citep{kikas2017structure} identify that utility packages are highly
central in package dependency networks with the power to affect more than 30\%
of all packages in the studied repositories.

In a comprehensive study of the evolution of package
repositories,~\cite{decan2019empirical} observe that three out of seven studied
repositories have superlinear growth of transitive relationships, forming and
strengthening new network hubs over time. Half of the packages in \crates, \npm,
and, NuGet had in 2017 at least $41$, $21$, and $27$ transitive dependencies,
nearly two times more than their respective number in 2015.
Although~\cite{decan2019empirical} finds that the number of dependencies a
developer declares in an application remains stable over time, the increasing
number of transitive relationships in package repositories is still an active
phenomenon after the \texttt{left-pad} incident. Apart from understanding the
structure and evolution of package repositories, researchers have also studied
known security vulnerabilities~\citep{decan2018impact,zimmermann2019small},
maintainability~\citep{valiev2018ecosystem,cogo2019empirical,
zerouali2018empirical}, software
reuse~\citep{abdalkareem2019impact,abdalkareem2017developers}, and more recently
breaking changes~\citep{mezzetti2018type,Mujahid_MSR2020} from a network
perspective.~\cite{zimmermann2019small} report that 40\% of \npm include a
package with a known vulnerability, suggesting that \npm forms a large attack
surface for hackers to exploit. Despite developer awareness on using trivial and
simple packages after the \texttt{left-pad}
incident,~\cite{abdalkareem2019impact} still find a prevalent number of
applications depending on trivial packages: 10\% of \npm and 6\% of PyPI
applications on \github depends on at least one package with less than 35 LOC.

Network analysis of packages commonly makes use of metadata from package manifests
to calculate the impact and severity of measured
variables.~\cite{ponta2018beyond}'s work on building a security dependency
checker using call graphs highlights the limitations of using metadata and the
importance of studying package dependencies with a contextual lens. Typically,
a subset of an API is vulnerable---not the entire package---and how clients
interact with API's is also highly contextual.~\cite{zapata2018towards} observed
through manual analysis that 75\% of 60 warned JavaScript projects did not
invoke the vulnerability. As an alternative to vulnerability detection through
call graphs,~\cite{chinthanet2020code} explores the idea of building
hierarchical structures of applications and their dependencies for Node.js. To pitch for code-centric instead of metadata-based
representations of package repositories,~\cite{hejderup2018software} propose
dependency networks based on function calls which we concretize in this work.

By embedding function call relationships into package dependency networks, we
aim to also bridge the gap between API and package repository research. Notably,
\ename could resolve the limitation of studying immediate API calls to include
chains of API calls (i.e., transitive calls) such as
in~\cite{robbes2012developers}'s work on determining the ripple effects on
deprecated APIs in the Smalltalk ecosystem. Similarly, combining qualitative
studies such as looking into
deprecation~\citep{sawant2018reaction,sawant2018understanding},
breakages~\citep{raemaekers2017semantic,xavier2017historical,bogart2016break},
and migration patterns~\citep{zhong2010mining,nguyen2019graph} with network
analyses could provide an additional empirical dimension in such studies. In
support of this,~\cite{zhangenabling}'s need-finding study calls for tooling that
supports API designers with data-driven recommendations, for example, on when to deprecate an
API.

\begin{figure}[tb]
   \centering
   \subfloat[State of the art: Package-based Dependency Networks.]{\label{fig:dep_nw_package}\includegraphics[width=0.8\columnwidth]{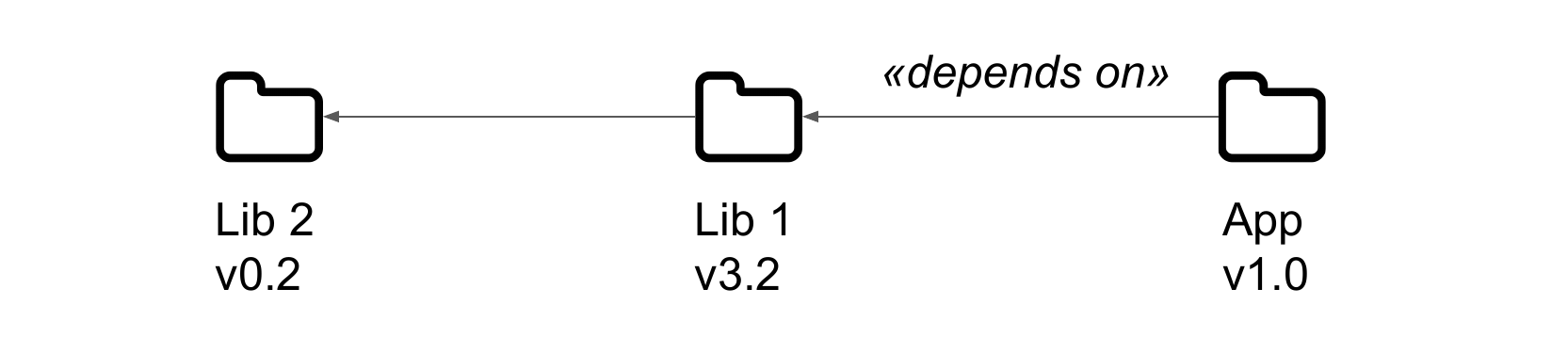}}\\
   \subfloat[Our proposal: Call-based Dependency Networks.]{\label{fig:dep_nw_call}\includegraphics[width=0.8\columnwidth]{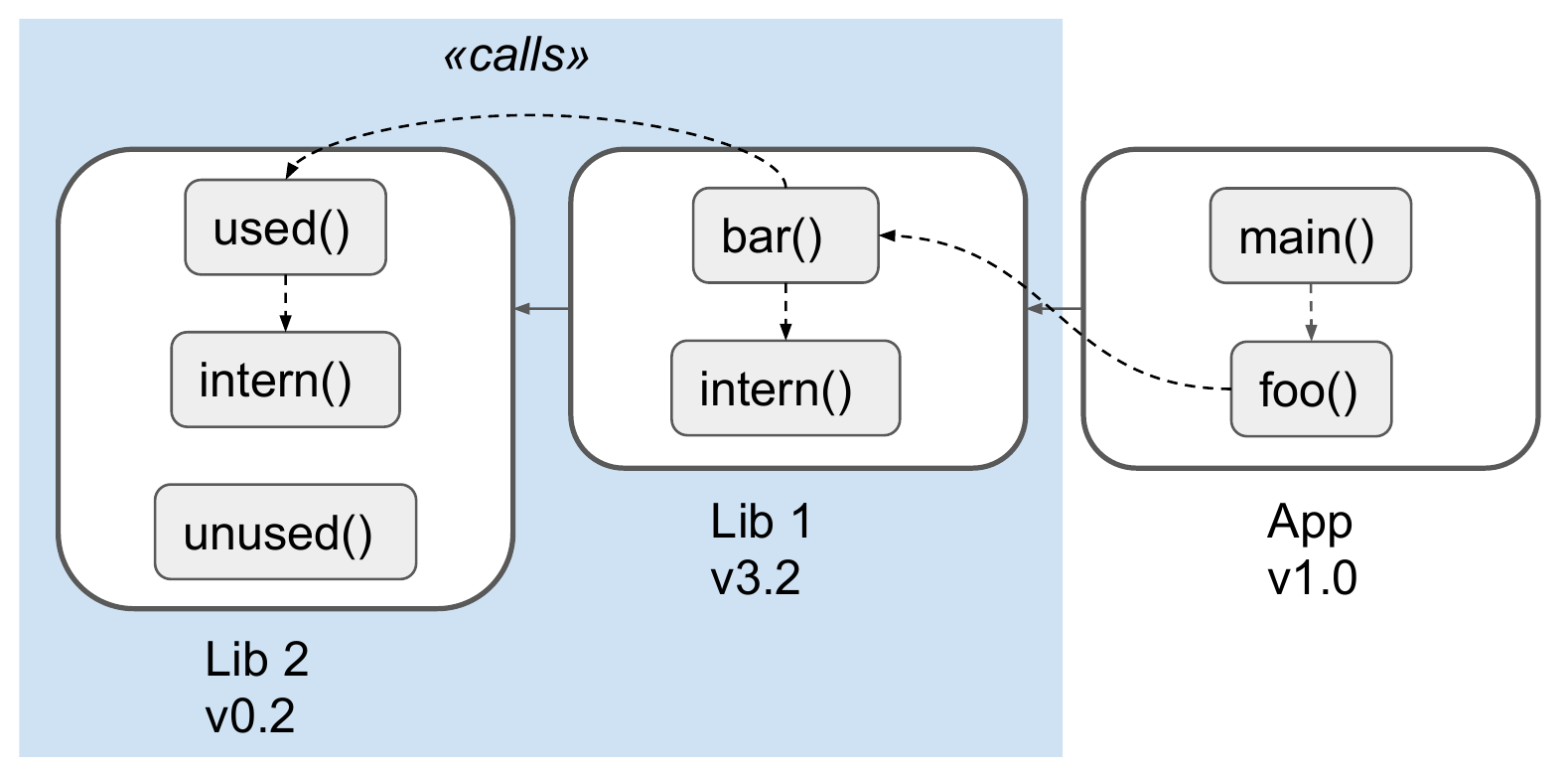}}\\
   \caption{Different granularities of dependency networks.}
   \label{fig:dep:nw}
\end{figure}

\subsection{Rust Programming Language}
Rust is a relatively new (first stable release 1.0 in
2015)\footnote{\ahref{http://web.archive.org/web/20180416152826/https://blog.rust-lang.org/2015/05/15/Rust-1.0.html}{https://blog.rust-lang.org/2015/05/15/Rust-1.0.html}}
systems programming language that aims to combine the speed of C with the memory
safety guarantees of a garbage-collected language such as Java. Rust is also
unique because its package management system (\cargo) was designed from the
ground-up to be part of the language environment~\citep{blogrustcargo}. \cargo
not only manages dependencies but prescribes a build process and a standardized
repository layout which helps facilitate the creation of automated large-scale
analyses such as \ename. Every \cargo package contains a file called
\texttt{Cargo.toml}, specifying dependencies on external packages. Moreover,
with \crates, there is one central place where all Rust packages (so-called
``crates'') live. As of 13 August 2020, \crates is the fifth most fast-growing
package repository hosting over $44,745$ packages and averaging 60 new packages
per
day.\footnote{\ahref{http://web.archive.org/web/20201201224023/http://www.modulecounts.com/}{http://www.modulecounts.com/}}

\subsection{Call-based Dependency Networks}\label{sec:cdn}

We distinguish two kinds of dependency networks, shown in \Cref{fig:dep:nw}:
i)~coarse-grained \emph{Package-based Dependency Networks} shown in
\Cref{fig:dep_nw_package}, similar to what dependency resolution tools (e.g.,
\cargo or \maven) build internally or what researchers have used in the past,
and ii)~fine-grained \emph{Call-based Dependency Networks} shown in
\Cref{fig:dep_nw_call}, which we advocate in this paper.

\Cref{fig:dep_nw_package} models an example of an end user application
\texttt{App}, which directly depends on \texttt{Lib1} and transitively depends
on \texttt{Lib2}. In such a PDN, each node represents a versioned package. An
edge connecting two nodes denotes that one package imports the other, for
example \texttt{App 1.0} depends on \texttt{Lib1 3.2}.

\Cref{fig:dep_nw_call} consists of three individual call graphs for
\texttt{App}, \texttt{Lib1}, and \texttt{Lib2}. Each of these call graphs
approximate internal function calls in a single package. Every node represents
a function by its name. The edges approximate the calling relationship between
functions, e.g., from \texttt{main()} to \texttt{foo()} within \texttt{App} in
\Cref{fig:dep_nw_call}. However, the function identifiers bear no version, nor
do they have globally unique identifiers (e.g., there are two \texttt{intern()}
functions in \Cref{fig:dep_nw_call}). We merge these two graph representations
to produce a CDN:

\begin{defi}{Definition}
  A \textbf{Call-based Dependency Network} (CDN) is a directed graph $G = \langle V, E
  \rangle$ where:
  \begin{enumerate}
    \item $V$ is a set of versioned functions. Each $v \in V$ is a tuple
      $\langle \mbox{\texttt{id,ver}} \rangle$, where \texttt{id} is a
      unique function identifier and \texttt{ver} is a float
      value depicting the version of the package in which \texttt{id} resides.
    \item $E$ is a set of edges that connect functions. Each $\langle
      \mbox{\texttt{v$_1$,v$_2$}} \rangle \in E$ represents a function call from
      \texttt{v}$_1$ to \texttt{v}$_2$.
  \end{enumerate}
\end{defi}

Applying the above definitions, the function \texttt{used()} in
\Cref{fig:dep_nw_call} becomes a node with the fully qualified identifier
\mbox{$\langle \mbox{\texttt{Lib2::used, 0.2}} \rangle$} $\in V$. The dependency
between \texttt{App} and \texttt{Lib1} is represented as $\langle$$\langle
\mbox{\texttt{App::foo, 0.1}}\rangle$, $\langle \mbox{\texttt{Lib1::bar,
3.2}}\rangle$$\rangle$ $\in E$.

CDNs offer a white-box view of the more coarse-grained PDNs. In particular, we can see that \texttt{unused()} is never called.  If
\texttt{unused()} was affected by a vulnerability, we can deduce from
\Cref{fig:dep_nw_call} that we should \emph{not} issue a security warning for
\texttt{App}, since it does not use the affected functionality. In contrast to
the CDN, the PDN in \Cref{fig:dep:nw} by its nature cannot provide such a
fine-grained precision level.

\section{\ename: Generating CDNs from Package Repositories}
In this section, we describe a generic approach, \ename, to systematically
construct CDNs for package repositories. \ename can be
applied to any programming environment that features i)~a way of expressing
dependency information between packages, and ii)~tooling to generate call graphs
for a package.

\begin{figure*}[tb]
   \includegraphics[width=1\columnwidth,keepaspectratio]{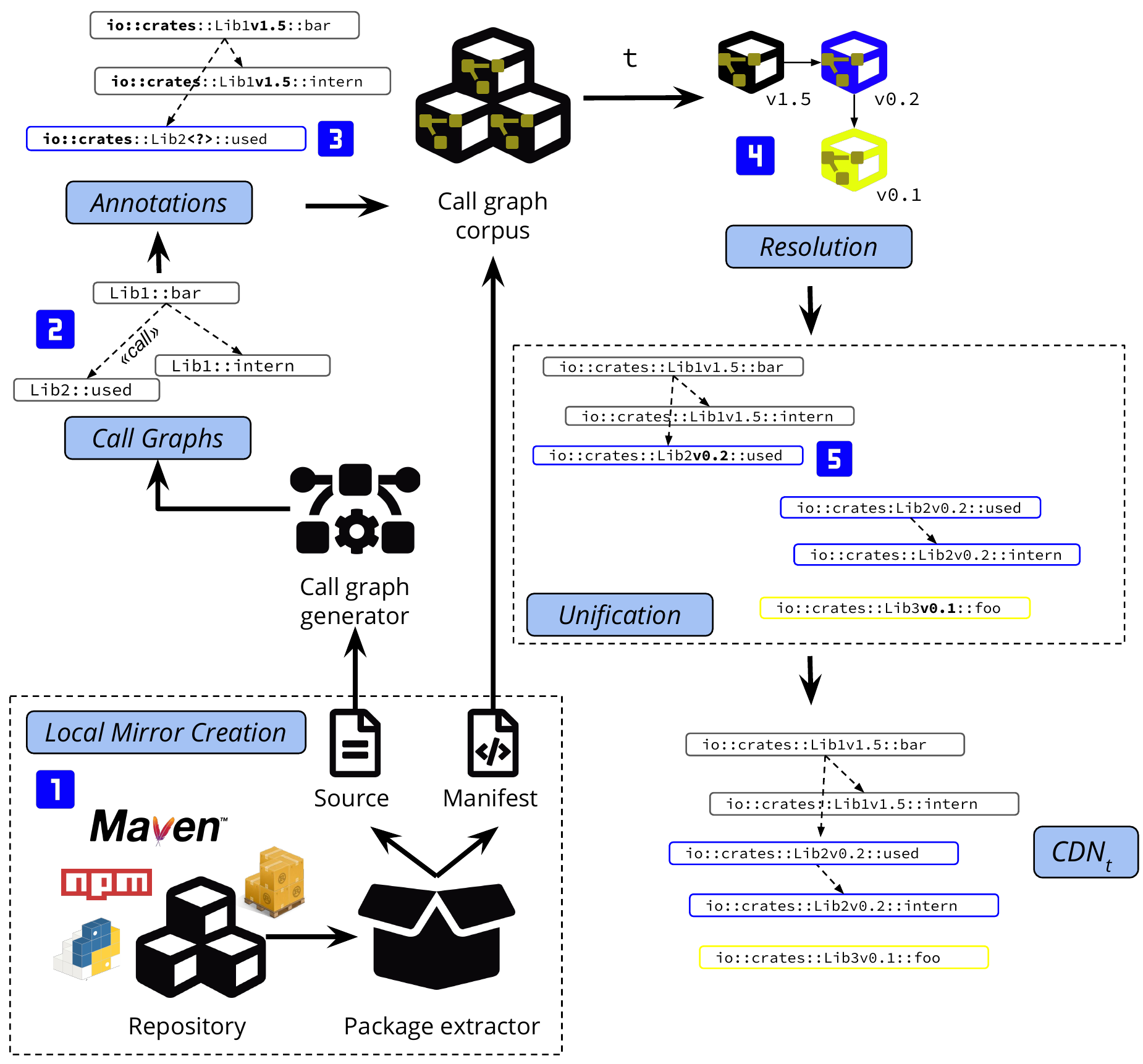}
   \caption{Generic approach to generate CDNs from package repositories}
   \label{fig:approach}
\end{figure*}

\ename constructs a CDN in a two-phase process illustrated in
\Cref{fig:approach}. In the first phase, \textit{Call Graph Generation} (step
\rectangled{1}, \rectangled{2}, and \rectangled{3}), \ename generates
a static dataset of annotated call graphs from packages in a repository.
In the second phase, \textit{Temporal Network Generation} (step \rectangled{4}
and \rectangled{5}), \ename first generates an intermediate package dependency
network by resolving dependencies between packages at a user-provided timestamp
$t$, and then unifies the call graphs of resolved packages into one temporal
call-based network, the $CDN_{t}$.

\subsection{Call Graph Generation}

\paragraph{\textbf{Local Mirror}}
Package managers keep an updatable index of package repositories to lookup
available packages and their versions. \ename uses such indices to extract
and download available packages in step \rectangled{1} in order to create local mirrors
of repositories (i.e., clones of repositories). A minimal local mirror needs to
contain the manifest and publication (or creation) timestamp for each version of
a package.

\paragraph{\textbf{Package Call Graphs}}

A call graph is a data representation of relationships between functions in a
program and serves as a high-level approximation of its runtime
behavior~\citep{ryder1979constructing,ali2012application}. From a static analysis
perspective, a call graph is useful for investigating and understanding
interprocedural communication between code elements (i.e., how functions
exchange information). In \ename, we view a call graph as a partial graph of a resulting CDN. We increase the scope of a call graph
from a single package (i.e., program) to a package and its dependencies. We
denote inter-package function relationships as the actual specific code
resources that packages use between each other (i.e., a dependency relationship
at the function granularity) and are first-class citizens in CDN analyses. The
call from \texttt{Lib1::bar} to \texttt{Lib2::used} in \rectangled{2} exemplifies
an inter-package function relationship. \ename requires nodes in call
graphs to have function identifiers with fully resolved return types and
arguments.

In the presence of dynamic features, such as virtual dispatch or reflection, there are implications to the precision and soundness of
call graphs that indirectly also affect generated CDNs.
Theoretically, it is impossible to have both a precise and sound call graph of a
program. Thus, \ename uses \textit{soundy}
call graph algorithms that follow a best-effort approach for the resolution of most language
features~\citep{livshits2015defense}. Precise yet unsound call graph algorithms could miss actual
inter-package function calls, making certain dependency analyses (e.g.,
security) of CDNs incomplete. Examples of soundy call graph algorithms for typed
languages include subclasses of Class Hierarchy Analysis
(CHA)~\citep{tip2000scalable,sundaresan2000practical} and Points-to
analyses~\citep{steensgaard1996points,shapiro1997fast,emami1994context} such as
$k$-CFA~\citep{shivers1991control}. In the case of untyped languages such as
Python or JavaScript, a
middle-ground is hybrid approaches combining both dynamic analysis and static
analysis such as~\cite{alimadadi2015hybrid}'s Tochal or~\cite{salis2021pycg}'s PyCG.

\paragraph{\textbf{Annotating Call Graphs}}\label{cg:annot}

To prepare call graphs for unification, we
need to rewrite function identifiers in each package call graph so that they
are globally unique. Without annotating function identifiers, inconsistencies
can arise from packages that have identical namespaces and multiple versions of
the same package in a dependency tree. \ename solves these issues by annotating
the function names, return types, and argument types in function signatures with
three components: i) repository name, ii) package name, and iii) static or
dynamic (i.e., constraint) package version.

For each function signature in a call graph of a package version, \ename maps
each type identifier found in the signature to the package that declares it. There are
three potential mappings of a type identifier to a package that do not reside in the standard
library of the language:

\begin{itemize}
   \item~\textbf{Local}, resulting in an annotated qualifier with the repository
   name, and its package name and version as exemplified in
   \texttt{io::crates::Lib1v1.5::bar}.
   \item~\textbf{Dependency with a static version}, resulting in an annotated qualifier
   with the repository name, and the name and version of the dependency.
   \item~\textbf{Dependency with a dynamic version}, resulting in an annotated
   qualifier with the repository name and name of the dependency. However the
   version is missing as exemplified in \texttt{io::crates::Lib2<?>::used} in \rectangled{3}.
\end{itemize}

The first two mappings denote a resolved type annotation, and the last one is an
unresolved type annotation. Function identifiers
with unresolved type annotations have their dynamic versions resolved to a
specific version at dependency resolution time (i.e., at the \textit{Temporal
Network Generation} phase). Finally, \ename splits the annotated call graph into two
sections, one immutable section containing resolved function
signatures, and another section containing unresolved function signatures. The
annotated call graphs are then stored in a dataset. The final dataset should
contain all downloaded packages that include creation timestamp, manifest file,
and annotated call graph with global identifiers. 

\subsection{Temporal Network Generation}\label{sec:tng}

\begin{figure}[tb]
   \centering
   \subfloat[Package A depends on B version \textit{1.*}.]{\label{fig:retrodepres_1}\includegraphics[width=0.5\columnwidth]{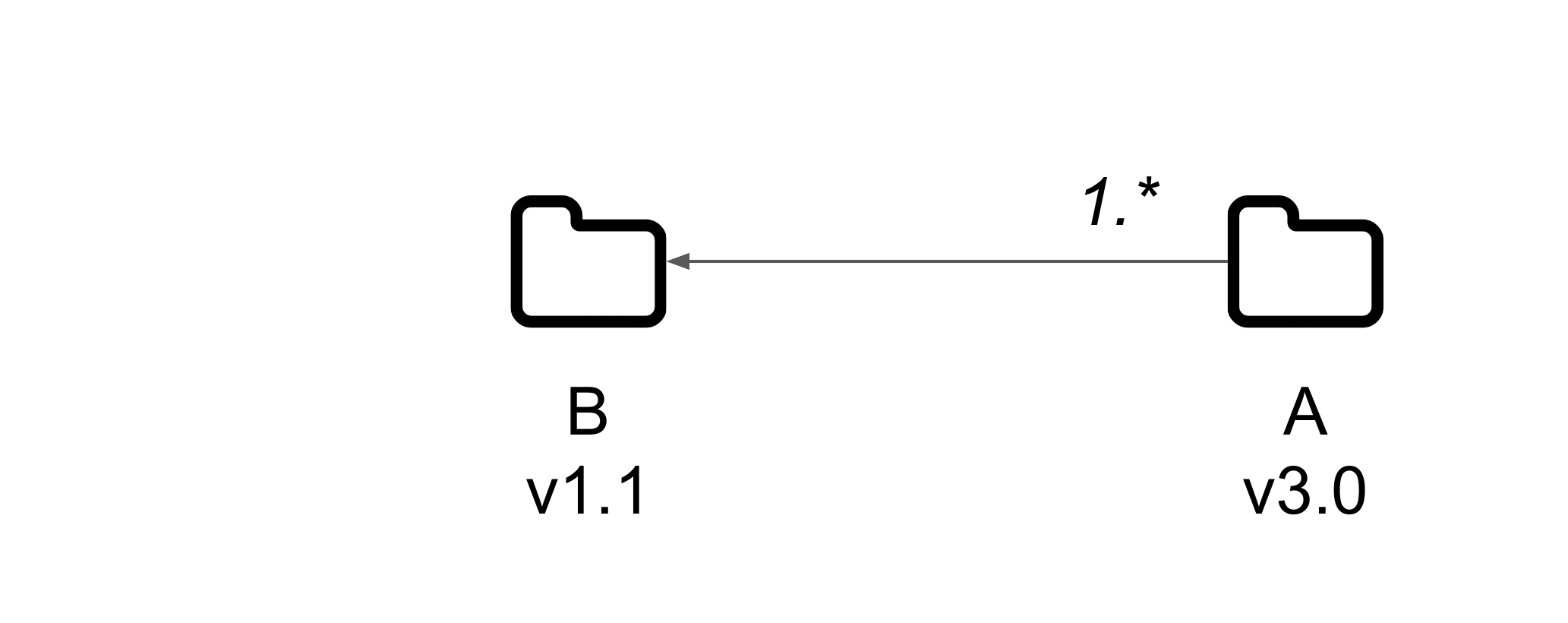}}
   \hfill
   \subfloat[Full dependency resolution tree with time.]{\label{fig:retrodepres_2}\includegraphics[width=0.5\columnwidth]{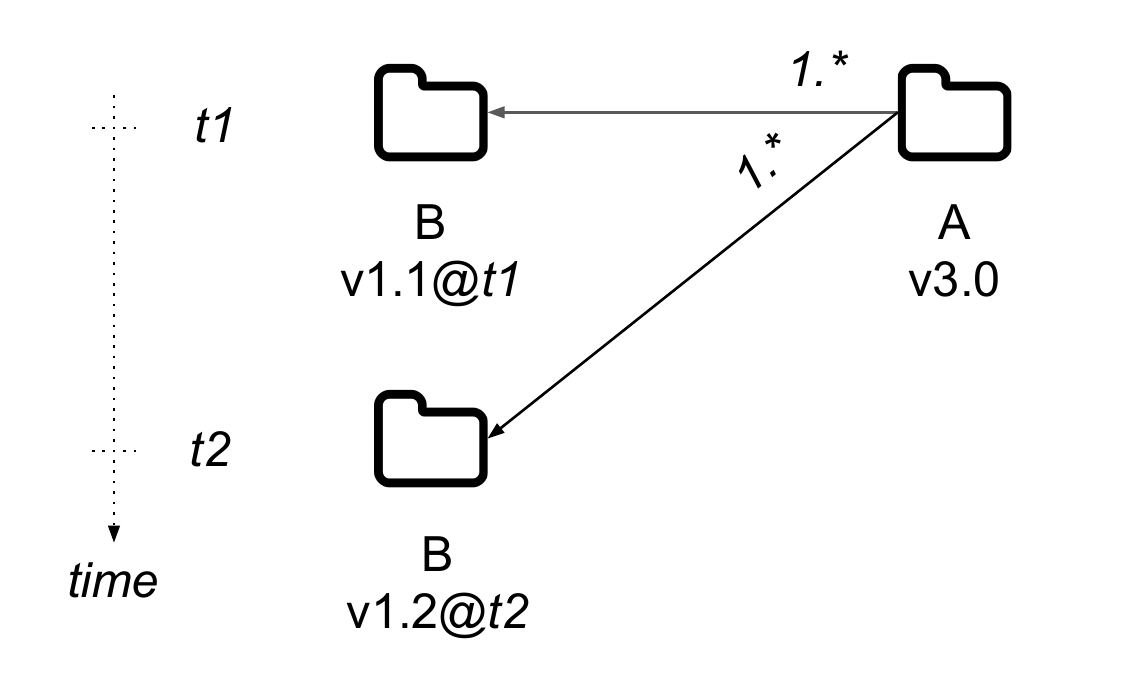}}
   \caption{Retro-active dependency resolution}
   \label{fig:retrodepres}
\end{figure}

\paragraph{\textbf{Retro-active Dependency Resolution}} 
To study the evolution of the relationships between packages in a repository, we
perform retroactive dependency resolution \rectangled{4} that generates a
concrete dependency network valid at a given timestamp $t$. The use of dynamic
versions in package manifests complicates network generation of package
repositories. During resolution time of package dependencies, a dynamic version
instructs the dependency resolver  to fetch the
most recent version within its allowed version boundary, making the relationship
between packages contemporary. Package $A$ depending on the dynamic version
$1.*$ of package $B$ that satisfies any version with a leading \emph{1.} (e.g.,
\emph{1.0},\emph{1.8}, or \emph{1.20.2}) in \Cref{fig:retrodepres_1} exemplifies
a dynamic version. Given that Package $B$ releases version $1.1$ at $t_1$ and
$1.2$ at $t_2$ ($t_1 < t_2$) in \Cref{fig:retrodepres_2}. At $t$, where $t_1 < t
< t_2$, a dependency resolver will select version $1.1$. However, at $t > t_2$,
it will select version $1.2$, highlighting the temporal changes in package
relationships. 

Given a timestamp $t$, \ename creates a subset $mirror_t$ of our local mirror
(i.e., copy of the \crates index)
containing packages and versions with a creation timestamp $t_c$ satisfying $t_c
\leq t$. Then, for each package version manifest file in $mirror_t$, we resolve its
dependencies using a dependency resolver. Dependency resolvers are usually
integrated into package managers and are available as independent libraries. 

\vspace{-0.2em}
\paragraph{\textbf{Call Graph Unification}} The unification is a two-phase
process. In the first phase, we build a resolved dependency tree for each
package version in $mirror_t$ and then perform a level-order traversal of each tree
to merge call graphs of child nodes with their parent nodes. The output is a
unified call graph of statically dispatched function calls for each
package version in $mirror_t$. In the merge phase of a parent and a child call graph,
we complete the unresolved function identifiers in the parent call graph with
the resolved version available in the child node. The function
\texttt{io::crates::Lib2v0.2::used} in \rectangled{5} replaces the unresolved
function \texttt{io::crates::Lib2<?>::used} in \rectangled{3} with $v0.2$.

In the second phase, we need to deal with dynamically dispatched functions and
localize call targets across package boundaries. To illustrate this process, we
introduce the following scenario: Package \texttt{A} depends on package
\texttt{B} and package \texttt{C}. Both \texttt{B} and \texttt{C} depend on the
library \texttt{serde}. Furthermore, \texttt{B} has a class
\texttt{Foo} that implements the function \texttt{serialize()} in the
\texttt{Serialize} interface of \texttt{serde}. \texttt{C} has a function called
\texttt{bar()} that takes a \texttt{Serialize}-like object as an argument and
invokes the dynamically-dispatched \texttt{serialize()} call on the object. 

Before merging the call graphs (i.e. first phase), \texttt{bar()} is only aware
of call targets that are within \texttt{C}. In this example, there are no call
targets available (i.e., there is no function implementing \texttt{serialize()}
in \texttt{C}). Thus, in the second phase, we search for other compatible
function implementations across packages that are available after merging their
call graphs. Here, we would create a call target from \texttt{bar()} in
\texttt{C} to the \texttt{serialize()} implementation in \texttt{Foo} in
\texttt{B}. It is possible that \texttt{A} may never pass an object of
\texttt{Foo} from \texttt{B} to function \texttt{foo()} in \texttt{C} in
practice. However, the second phase is necessary to ensure that dynamically
dispatched functions remain sound after merging all call graphs together.

After constructing a package-level call graph for each package version in $mirror_t$,
we merge all partial call graphs into a single CDN. The process consists of
aggregating all package-level call graphs and then merging them to remove
duplicate nodes and edges. The result is a CDN corresponding to the package
repository at the given timestamp $t$.

\section{Implementing \ename for \crates}\label{sec:cdn-impl:crates} 
We implement \ename for \crates, the official package repository for
Rust. Unlike mainstream package repositories such as \mavencentral, PyPI, \npm,
and NuGet, \crates do not host pre-built binaries but the source code of its
packages. To generate call graphs for Rust packages, we need to first perform a
large-scale compilation of \crates and then extract call graphs from generated
binaries. Attempting to reproduce the build of a piece of software is known to
be challenging~\citep{sulir2016quantitative},~\cite{tufano2017there}'s compilation of $219,395$ Apache snapshots yielded
a success rate of 38\%, and~\cite{martins201850k}'s compilation of $353,709$ Github
Java projects yielded a success rate of 56\%. An overall low success rate could
potentially endanger representative studies of \crates.

In the remainder of this section, we describe key implementation choices and
results from our large-scale compilation of \crates. 

\paragraph{Creating a local mirror}
We clone a snapshot of \crates's official git-based
index\footnote{\ahref{https://web.archive.org/web/20180224105846/https://github.com/rust-lang/crates.io-index}{https://github.com/rust-lang/crates.io-index}}
at revision \texttt{6c550c8} (14th February 2020) containing $35,896$ packages,
$208,023$ releases, and $1,151,001$ dependency relationships. By validating the
dependency specification in the index for invalid names or dependency
constraints, we can save resources by avoiding building broken releases. We
identify $1,506$ releases from $201$ packages having dependencies that do not
match existing packages, and $5,667$ releases from $4,427$ packages having
dependencies with unsolvable constraints (i.e., no available versions for the
constraint)

The documentation hosting service for \crates,
\texttt{Docs.rs},\footnote{\ahref{http://web.archive.org/web/20201201224058/https://github.com/rust-lang/docs.rs}{https://github.com/rust-lang/docs.rs}}
provides Rust users API documentation for every published package release. In
addition to automatically generating documentation for package releases,
\texttt{Docs.rs} also documents the build log and compile status publically. We
create a web scraper that extracts the build status on \texttt{Docs.rs} for each
release in our dataset. In total, we found that $43,893$ indexed releases
belonging to $10,154$ packages have build failures, amounting to 20\% of
\crates. In addition to the \crates index, we use \texttt{Docs.rs} as externally validated metadata source in our study. 

After subtracting build failures and invalid dependency specifications, our
final index amounts to $156,484$ releases from $29,480$ packages. Lastly, we use
the official API at
\ahref{http://web.archive.org/web/20201201224106/https://crates.io/api/v1/crates}{https://crates.io/api/v1/crates}
to download all packages and their creation timestamp (not available in the index).

\begin{sloppypar}
\paragraph{Choosing a Call Graph Generator}
There are two approaches for constructing a call graph from a Rust program, the
higher-level LLVM
analysis,\footnote{\ahref{https://web.archive.org/web/20180517123938/https://llvm.org/docs/Passes.html}{https://llvm.org/docs/Passes.html}}
and the lower-level MIR analysis~\citep{rustblog:mir}. Rust functions and its
calls are either of monomorphized (i.e., static dispatch) or virtualized (i.e.,
dynamic dispatch) nature. From the
documentation\footnote{\ahref{http://web.archive.org/web/20201201224110/https://doc.rust-lang.org/book/ch03-03-how-functions-work.html}{https://doc.rust-lang.org/book/ch03-03-how-functions-work.html}}
and a comprehensive benchmark~\citep{rustforum:cgbench}, we can learn that there
are two monomorphized features,
\texttt{macros}\footnote{\ahref{http://web.archive.org/web/20201201220221/https://doc.rust-lang.org/stable/reference/macros.html}{https://doc.rust-lang.org/stable/reference/macros.html\#trait-objects}}
and \texttt{generic functions}, and two virtualized features, \texttt{trait
Objects},\footnote{\ahref{http://web.archive.org/web/20201201220211/https://doc.rust-lang.org/stable/reference/types.html}{https://doc.rust-lang.org/stable/reference/types.html}}
and \texttt{function
pointers},\footnote{\ahref{http://web.archive.org/web/20201201220224/https://doc.rust-lang.org/book/ch19-05-advanced-functions-and-closures.html}{https://doc.rust-lang.org/book/ch19-05-advanced-functions-and-closures.html}}
that dispatch functions in Rust.

As part of the output in compilation of Rust programs, we can use the optionally generated LLVM
IRs\footnote{\ahref{http://web.archive.org/web/20201201220255/https://doc.rust-lang.org/rustc/command-line-arguments.html}{https://doc.rust-lang.org/rustc/command-line-arguments.html\#--emit-specifies-the-types-of-output-files-to-generate}}
to build call graphs using the LLVM call graph
generator\footnote{\ahref{http://web.archive.org/web/20201201220252/http://llvm.org/doxygen/CallGraph_8h_source.html}{http://llvm.org/doxygen/CallGraph_8h_source.html}} or
\texttt{cargo-call-stack}~\citep{cargo:call:stack}. Due to the absence of
Rust-specific type information in LLVM
IRs,\footnote{\ahref{http://web.archive.org/web/20201201224720/https://github.com/rust-lang/rust/issues/59412}{https://github.com/rust-lang/rust/issues/59412}} call graph
generators can only resolve monomorphized features and cannot provide complete
type information needed in \ename. By analyzing at the MIR level instead of the
LLVM level,~\texttt{rust-callgraphs}\footnote{\url{https://github.com/ktrianta/rust-callgraphs}} offers a more feature-complete
call graph by implementing a CHA algorithm and is our choice for building CDNs. In addition to monomorphized features, it can resolve function calls
dispatched through \texttt{Trait Objects}, making it a more soundy choice over
the LLVM-based call graph generators. Although \texttt{rust-callgraphs} does not
support \texttt{function pointers}, it is a negligible trade-off as the
documentation\footnote{\ahref{http://web.archive.org/web/20201201220303/https://rust-lang.github.io/unsafe-code-guidelines/layout/function-pointers.html}{https://rust-lang.github.io/unsafe-code-guidelines/layout/function-pointers.html}}
state that \texttt{function pointers} are mostly useful for calling C code from
Rust.

For annotating call graphs, the metadata in call graph nodes contains
package information and access identifiers. Moreover, the complementary type
hierarchy output contains complete type information for creating resolved
function identifiers. We also keep the edge metadata that includes dispatch
information (i.e., static, dynamic, or macro) in the annotated call graphs.
\end{sloppypar}

\begin{table}[tb]
\caption{Build statistics}
\label{tab:buildstats}
\centering
\begin{adjustbox}{width={0.9\columnwidth},totalheight={0.9\textheight},keepaspectratio}
\begin{tabular}{@{}lrrr@{}}
\toprule
{\bfseries Build} & {\bfseries \#Releases} & {\bfseries \#Packages} &
{\bfseries Time (hrs)} \\
\midrule
\crates index & $208,023$ & $35,896$  & --- \\
\midrule
\texttt{Docs.rs} & $156,484$ (-24.78\%) & $29,480$ (-17.87\%)  & --- \\
\texttt{rust-callgraphs} & $142,301$(-9.06\%) & $23,767$ (-19.38\%)  & 10 days \\
\bottomrule
\end{tabular}
\end{adjustbox}
\end{table}

\paragraph{Large-Scale Compilation of \crates}
Some Rust packages depend on external system libraries such as
\texttt{libavcodec} or \texttt{libxml2} to successfully compile. Knowing which
external libraries to install for compiling such packages is a manual and
tedious process. Luckily, the Rust infrastructure team maintains a Docker image,
\ahref{http://web.archive.org/web/20201201224730/https://github.com/rust-lang/crates-build-env}{rust-lang/crates-build-env},
that bootstraps a Rust build environment pre-installed with community curated
systems libraries, increasing the chances for successful compilations. We use
\ahref{http://web.archive.org/web/20201201224727/https://github.com/rust-lang/rustwide}{Rustwide}, an API for spawning Rust
build containers, and configure it to use \texttt{rustc 1.42.0-nightly} compiler
together with \texttt{rust-callgraphs}'s compiler plugin. After compilation, we
use the analyzer component in \texttt{rust-callgraphs} to generate and store the
call graphs in our dataset.

We set up a compilation pipeline on four build servers running 34 docker
containers to compile packages and build call graphs. It took 10 days to
complete it. \Cref{tab:buildstats} shows the compilation results in comparison
with index entries and \texttt{Docs.rs} results. Overall, our call graph corpus (CG
Corpus) has a call graph for 90\% of all compilable versions (70\% of all
indexed versions) and at least one version for 80\% of all packages built by
\texttt{Docs.rs}. The high success rate showcases the practical feasibility of
\ename for \crates.

\begin{sloppypar}

\paragraph{Dependency Resolution}
For each \cargotoml manifest in our downloaded dataset, we extract dependencies
intended for source code use. These include library dependencies (i.e.,
\texttt{[dependencies]}), platform-specific dependencies (i.e.,
\texttt{[target]}), and also enabled optional dependencies in
\texttt{[features]}. Both~\cite{kikas2017structure}
and~\cite{decan2019empirical} do not take into account both enabled optional
dependencies and platform-specific dependencies, considering only library
dependencies when analyzing \crates.
\end{sloppypar}
The \cargotoml manifest supports specifications of dependencies using the
\texttt{semver} schema~\citep{semver}. A version is a three-part version number:
major version, minor version, and patch version. An example of a version is
$1.0.0$. An increase in the major number denotes incompatible changes, an
increase in the minor number denotes backward-compatible changes, and an
increase in the patch number denotes small bug fixes. With the support of range
operators (i.e., dynamic version) in dependency specifications such as caret
(e.g., $\wedge1.0.0$), tilde (i.e., $\sim1.0.0$), wildcard (e.g., $1.*$), and
ranges (e.g., $>1.0.0. <= 2.0.0$), the dependency resolver in \cargo will
attempt to resolve the latest version satisfying the constraint. When multiple
constraints of the same dependency appear in the dependency tree, \cargo first
attempts to find the most recent version satisfying all constraints. For
example, for the two constraints, \texttt{log} \textit{0.4.*} and \texttt{log} \textit{0.4.4}, the
dependency resolver will resolve \texttt{log} \textit{0.4.4}. However, for example, if
the resolver has to resolve \texttt{log} \textit{0.4.*} and \texttt{log} \textit{0.5.*}, there is
no single compatible version that matches both constraints. Instead, the
resolver will include two versions of the same dependency (e.g., \texttt{log}
\textit{0.4.4} and \texttt{log} \textit{0.5.5}) through name mangling to avoid conflicts~\citep{blogrustcargo}. The
resolution strategy of having multiple versions of the same dependency is
similar to
\npm.\footnote{\ahref{https://web.archive.org/web/20210426093903/http://npm.github.io/npm-like-im-5/npm3/dependency-resolution.html}{http://npm.github.io/npm-like-im-5/npm3/dependency-resolution.html}}

\begin{sloppypar} 
To emulate dependency resolution in Rust, we use the native Rust-implementation
of the \ahref{https://crates.io/crates/semver}{semver} library for use in Python
by invoking its native implementation through FFI (Foreign Function Interface) bindings.
Thus, we resolve dependencies and their constraints using the same library as
the \cargo package manager. For every timestamp $t$ in the CDN generation
process, we set the resolution to solve the latest version available at $t$ satisfying the constraint. 
\end{sloppypar}

\section{Structure and Evolution of the \crates CDN}
We address three research questions to contrast the similarities and differences
when using three different network sources (i.e., metadata, compile-validated
metadata, and control-flow data) for characterizing the structure and evolution
of \crates. In addition to comparing the networks, we also investigate how
reliably package-based dependency networks mirror the use of dependencies in the
source code.

\subsection{Research Questions}

\paragraph{\textbf{RQ1: What are the network characteristics of \crates?}}\mbox{}\\
\begin{sloppypar}
We characterize the calling relationship between packages in \crates, and then
identify various influential packages featuring a high number of callers and
callees within the networks. Specifically, we describe our data corpus and the
degree distribution to gain an overall understanding of the direct relationship
between functions for a large package repository such as \crates.

\end{sloppypar}
\paragraph{\textbf{RQ2: How does \crates evolve?}}\mbox{}\\
The frequent number of new package releases and the adoption of \texttt{semver}
range operators in dependency specifications make the relationship between
packages highly temporal in \crates. We capture these dynamics using both a
package-level perspective and the more fine-grained, function-level perspective.
In comparison to previous studies~\citep{kikas2017structure,decan2019empirical},
we use three different sources, namely metadata, compile-validated metadata, and
control-flow data, to understand their differences and similarities for
package-based dependency analysis. 

As all our snapshots deviate from a normal distribution according to
Shapiro-Wilk ($p < 0.01 \leq \alpha$), we use the non-parametric Spearman
correlation ($\rho$) coefficient for correlation analysis. Using Hopkins's
guidelines~\citep{hopkins1997new}, we interpret $0 \leq |\rho| < 0.3$ as no,
$0.3 \leq |\rho| < 0.5$ as a weak, $0.5 \leq |\rho| < 0.7$ as a moderate, and
$0.7 \leq |\rho| \leq 1$ as strong correlation We answer the following
\textbf{sub-RQs} using a package-level and call-level perspective:\\

\noindent
\textbf{RQ2.1: How do package dependencies and dependents evolve?}\\
\textbf{RQ2.2: How does the use of external APIs in packages evolve?}\\
\textbf{RQ2.3: How prevalent is function bloat in package dependencies?}\\
\textbf{RQ2.4: How fragile is \crates to function-level changes?}\\

For deciding on reasonable time points for evolution studies of package
repositories, we include a guideline with analysis in \Cref{apdix:verres}.

\paragraph{\textbf{RQ3: How reliable are dependency networks?}}\mbox{}\\

A dependency network approximates how packages use each other in a repository.
Both metadata-based networks and call-based networks have trade-offs and
limitations that affect how reliable they estimate actual package
relationships. To understand how accurate these networks are in practice, we
perform a manual analysis of 34 random cases where a metadata-based and call-based
dependency network infers relationships differently.  The cases involve both
direct and transitive package relationships.

\subsection{RQ1: Descriptive Analysis}

\begin{table}[tb]
\caption{Summary of Datasets}
\label{tab:ds}
\centering
\begin{tabular}{@{}l|r|r@{}}
\toprule
                              & \multicolumn{1}{l|}{\textbf{CG Corpus}} & \multicolumn{1}{l}{\textbf{CDN Feb'20}} \\ \midrule
\textbf{Functions}     & \textbf{121,825,729}           & \textbf{44,190,643}            \\ \midrule
~~~~~~~\ldots~public access                       & 46,236,696                     & 20,157,155                     \\
~~~~~~~\ldots~private access                      & 75,589,033                     & 24,033,488                     \\ \midrule
\textbf{Call edges}    & \textbf{327,535,934}           & \textbf{216,239,360}           \\ \midrule
\textbf{Intra-Package Calls} & \textbf{169,579,315}           & \textbf{102,136,956}           \\
~~~~~~~\ldots~macro invocation                        & 693,148                        & 356,329                        \\
~~~~~~~\ldots~static dispatch                       & 28,570,266                     & 20,650,000                     \\
~~~~~~~\ldots~dynamic dispatch                      & 140,315,901                    & 83,130,627                     \\
\textbf{Inter-Package Calls} & \textbf{157,956,619}           & \textbf{114,102,404}           \\
~~~~~~~\ldots~macro invocation                        & 7,183,797                      & 2,178,547                      \\
~~~~~~~\ldots~static dispatch                     & 29,650,173                     & 13,319,367                     \\
~~~~~~~\ldots~dynamic dispatch                      & 121,122, 649                   & 98,604,490                     \\ \bottomrule
\end{tabular}
\end{table}

\subsubsection{Summary of Datasets}
Before investigating the calling relationship among packages in \crates, we
first describe our datasets of generated call graphs (i.e., \textbf{CG Corpus})
and our largest CDN dated February 2020 in \Cref{tab:ds}. After removing all
function calls to the standard libraries of Rust, the call graph corpus has over
$121$ million functions and $327$ million function calls from $142,301$ compiled
releases. When merging call graphs into a CDN, we generate a compact
representation with over $44$ million functions and $216$ million function
calls, a sizeable reduction of $2.5$ and $1.5$ times of the \textbf{CG Corpus}
(i.e., functions and calls), respectively.

\Cref{tab:ds} also breaks down function calls into their dispatch type, namely
macro, static, and dynamic calls. Notably, nearly 80\% of all edges in the
\textbf{CG Corpus} are of a dynamic dispatch type, followed by static dispatch
(18\%) and macro invocations (2\%). The high number of dynamically dispatched
calls in the network indicates that \crates has a large pool of possible target
implementations to virtual functions---not necessarily magnitude more function
calls than statically dispatched calls. When comparing the access modifiers
between functions, we can see that 40\% of all functions inside \crates are
publicly consumable. Also, we can see that calling functions in
external packages is widespread in \crates; half of all the function calls
invoke a function from an external package (i.e., inter-package call). Unlike
the other two dispatch forms, 91\% of all macro dispatched calls exclusively
target macros defined in external packages. Overall, the high number of declared
public functions and the large degree of inter-package calls indicate that code
reuse in the form of functions between packages is a prevalent practice in
\crates.

\begin{mdframed}[roundcorner=0pt,nobreak=true,align=center]
Function reuse is prevalent; 40\% of functions are public and 49\% of call edges
target a dependency.
\end{mdframed}

\begin{figure*}[tb]
\includegraphics[width=1\columnwidth,keepaspectratio]{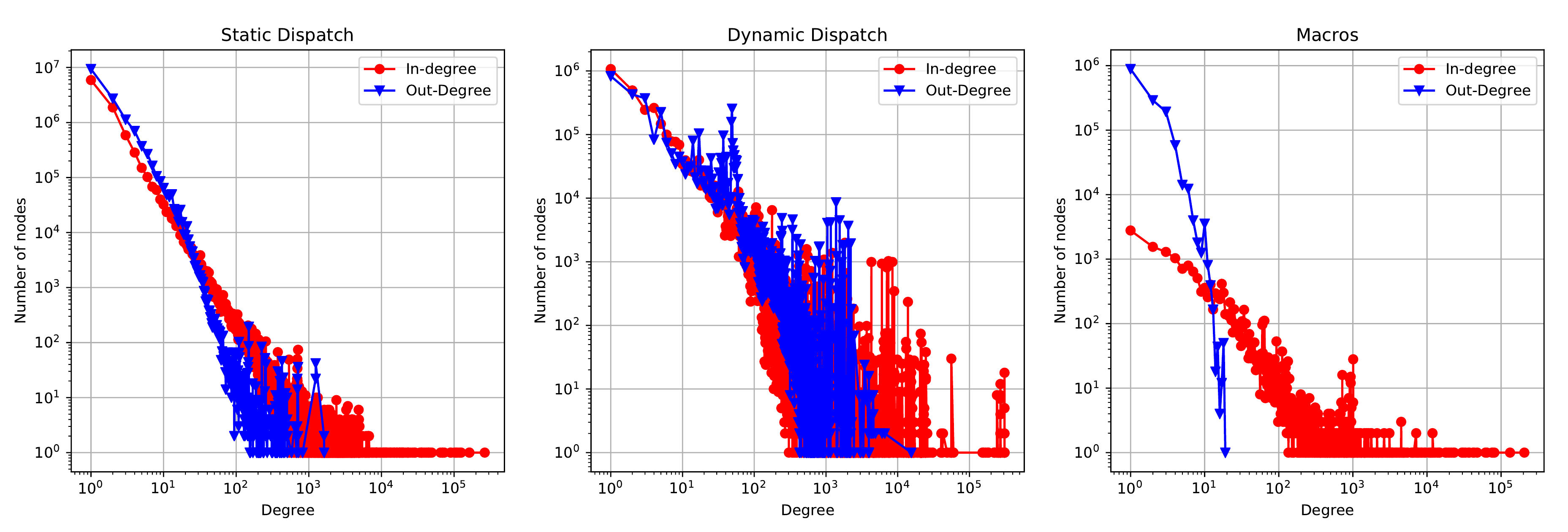}
\caption{Degree distribution of all function calls}
\label{fig:degree:overall}
\end{figure*}

\begin{figure*}[tb]
\includegraphics[width=1\columnwidth,keepaspectratio]{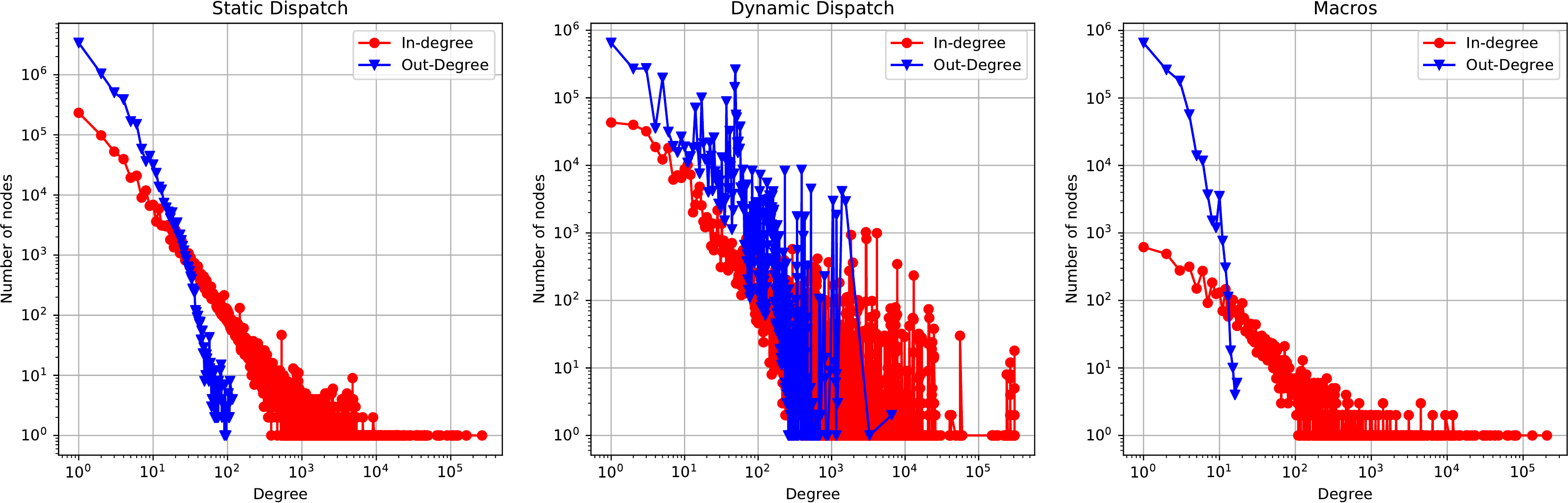}
\caption{Degree distribution of inter-package function calls}
\label{fig:degree:downstream}
\end{figure*}

\subsubsection{Function Call Distribution}\label{sec:desc:fns}
\Cref{fig:degree:overall} presents the degree distribution for all function calls
grouped by their dispatch type, and \Cref{fig:degree:downstream} is a
narrowed-down version looking at only inter-package function calls. The
out-degree of a function is the number of function calls to other unique
functions (i.e., number of caller-callee relationships). The in-degree of a
function is the number of callers to a function across \crates (i.e., number of
callee-caller relationships). Given a function \texttt{a()} in a package, the
out-degree looks at what calls \texttt{a()} makes. The in-degree looks at which
functions in \crates call \texttt{a()}. As mentioned earlier, inter-package calls 
are only function calls between
packages (i.e., pruning all internal calls). The
out-degree distribution for dynamic dispatch represents the number of possible
target functions in a virtual
method table,\footnote{\href{https://web.archive.org/web/20201112013908/https://alschwalm.com/blog/static/2017/03/07/exploring-dynamic-dispatch-in-rust/}{https://alschwalm.com/blog/static/2017/03/07/exploring-dynamic-dispatch-in-rust/}} and, for static- and macro dispatch, the number of function calls.
The in-degree distribution presents the aggregated number of callers for a
function (i.e., callee) and implementations of virtual functions for dynamic
dispatch, respectively. Overall, we can observe a long tail for both the
in-degree and the out-degree of
each dispatch mechanism, suggesting that the \crates CDN is a scale-free network
with the presence of a few nodes that are highly connected to other nodes in the
network (i.e., hubs). Finally,
\Cref{tab:static:degree,tab:cha:degree,tab:macro:degree} describe the top 5
functions with the highest in-degree and out-degree calls per dispatch type. The
top 5 list is an aggregation of functions per package. For example, the
\texttt{serde} package in \Cref{tab:cha:degree} has over 300 serialization
functions with an in-degree similar to $264,281$. Thus, we present the top 5
functions as the top most called function(s) per package. In
the following, we describe key results for each of the three dispatch forms.
\begin{sloppypar}

\begin{table}[tb]
\caption{The top 5 functions with most statically-dispatched calls}
\label{tab:static:degree}
\begin{adjustbox}{width={1\columnwidth},keepaspectratio}
\begin{tabular}{@{}llr|llr@{}}
\toprule
\multicolumn{3}{c|}{Outdegree}                   & \multicolumn{3}{c}{Indegree}          \\
Package              & Function          & \#    & Package     & Function       & \#      \\ \midrule
epoxy                & \texttt{load\_with}        & 1,625 & serde       & \texttt{missing\_field} & 264,281 \\
sv-parser-syntaxtree & \texttt{next, into\_iter}  & 1,243 & log         & \texttt{max\_level}     & 162,747 \\
python-syntax        & \texttt{\_\_reduce}        & 821   & vcell       & \texttt{set}            & 125,287 \\
rustpython-parser    & \texttt{\_\_reduce}        & 720   & serde\_json & \texttt{from\_str}      & 73,171  \\
mallumo-gls          & \texttt{load\_with}        & 712   & futures     & \texttt{and\_then}      & 65,043  \\ \bottomrule
\end{tabular}
\end{adjustbox}
\end{table}

\paragraph{Static dispatch} The median out-degree for statically dispatched
function call is 1 call (mean: 2.25) in both cases and at the 99th percentile
being 15 calls (13 calls for inter-package calls).  When comparing the
out-degree between statically dispatched calls in \Cref{fig:degree:overall} and
\Cref{fig:degree:downstream}, we can notice that there are 1865 functions
($0.012\%$) that call more than 100 other internal functions in
\Cref{fig:degree:overall}. The highest number of calls made by a single function
in both plots is to 1625 local functions and 116 external functions,
respectively. The relatively high number of internal function calls among the
outliers seems un-realistic at a first glance. Upon manual inspection of the
source code of the only two packages having functions with an out-degree greater
than 1000 (see \Cref{tab:static:degree}), namely
\texttt{epoxy}\footnote{\url{https://docs.rs/crate/epoxy/0.1.0/source/}} and
\texttt{sv-parser-syntaxtree}\footnote{\url{https://docs.rs/crate/sv-parser-syntaxtree/0.6.0/source/}},
we identify that this is the result of generic instantiations for creating
bindings to the \texttt{libepoxy} (an OpenGL function pointer manager) and
tokens for parsing SystemVerilog files.

The median in-degree for statically dispatched function calls are 1 (mean: 3.6)
and the 99th quantile is 24. When omitting all internal calls and considering
only inter-package calls, the median is 2 (mean: 24) and the 99th quantile is
208. There are three functions having over $100,000$ external calls in
\Cref{tab:static:degree}, \texttt{serde} for \textit{serialization},
\texttt{log} for \textit{logging}, and \texttt{vcell} for \textit{memory
management}. While the first two are the most downloaded and depended upon
packages in \crates, \texttt{vcell} stands out for only having nearly 300 dependent packages. After
inspection of the source code of those packages for the specific \texttt{set}
call, we could identify extensive implementations of low-level drivers to
interface various microcontrollers such as the \texttt{Cortex-M} and
\texttt{STM32} series.

\begin{table}[tb]
\caption{The top 5 functions with most dynamically-dispatched calls}
\label{tab:cha:degree}
\begin{adjustbox}{width={1\columnwidth},keepaspectratio}
\begin{tabular}{@{}llr|llr@{}}
\toprule
\multicolumn{3}{c|}{Outdegree}                          & \multicolumn{3}{c}{Indegree}                         \\
Package                 & Function             & \#     & Package           & Function                & \#      \\ \midrule
hyperbuild              & \texttt{match\_trie}          & 15,460 & serde             & \texttt{deserialize\_any} & 307,976 \\
heim-common             & \texttt{to\_base, from\_base} & 6,597  & serde\_json       & \texttt{from\_str} & 268,887 \\
uom                     & \texttt{to\_base, from\_base} & 6,045  & serde\_urlencoded & \texttt{deserialize\_identifier} & 59,737  \\
fpa                     & \texttt{I1F7, I2F6}           & 3,966  & yup-oauth2        & \texttt{token}                   & 42,737  \\
rtdlib                  & \texttt{deserialize}          & 2,470  & cpp\_core         & \texttt{cast\_into}              & 28,278  \\ \bottomrule
\end{tabular}
\end{adjustbox}
\end{table}

\paragraph{Dynamic dispatch}
We use \texttt{vtable} to refer to all
implementations of a virtual function of a Trait object. In practice, each Trait
object points to compatible Trait Implementations (having a \texttt{vtable} with
function and other member implementations). The median number of function targets function
\texttt{vtable} is 9 (mean: 42 (all), 32
(inter-package)) for both all function targets and only inter-package function
targets. The main deviation is at the 99th percentile, the outdegree for all
function targets is 800 for all targets, two times higher than when only considering inter-package
function targets. The highest out-degree function in \Cref{tab:cha:degree} is
\texttt{match\_trie} in the package \texttt{hyperbuild v0.0.10}, a HTML
minification library, having a \texttt{vtable} with $15,460$ function targets. The function
takes as an argument a \texttt{trie: \&dyn ITrieNode<V>} Trait, invoking
\texttt{get\_child} and \texttt{get\_value} of the Trait \texttt{ITrieNode}. The
Trait is implemented for all forms of HTML entities, explaining this high
outdegree value. In total, there are 38,352 ($< 0.94\%$) functions that populate
a vtable with more than 1000 function targets. Similarly, we can observe
$11,906$ ($< 0.36\%$) inter-package function calls with over $1000$ function targets.

The median in-degree for implementing a virtual (i.e. trait) function is 3
(mean: 53) and the 99th percentile is 608. When only considering inter-package
relationships, the median is 3 (mean: 64) and the 99th percentile is 875. As
shown  in \Cref{tab:cha:degree}, the most commonly implemented trait function
stems from
serializer packages such as \texttt{deserialize\_any} in \texttt{serde},
\texttt{from\_str} in \texttt{serde\_json} and \texttt{deserialize\_identifier}
in \texttt{toml}. In addition to serialization functions, we can also observe
that $42,737$ functions implement the trait function \texttt{token} in \texttt{yup-oauth2}
for user authentication with OAuth 2.0.  

\begin{table}[tb]
\caption{The top 5 functions with most macro-dispatched calls}
\label{tab:macro:degree}
\begin{adjustbox}{width={1\columnwidth},keepaspectratio}
\begin{tabular}{@{}llr|llr@{}}
\toprule
\multicolumn{3}{c|}{Outdegree}                     & \multicolumn{3}{c}{Indegree}                                  \\
Package              & Function               & \# & Package      & Function                              & \#      \\ \midrule
item                 & \texttt{path\_segment}          & 19 & log          & \texttt{log!}                                  & 205,810 \\
fungi-lang           & \texttt{fgi\_module}            & 18 & bitflags     & \texttt{\_\_impl\_bitflags!}                   & 77,848  \\
syn                  & \texttt{path\_segment}          & 17 & lazy\_static & \texttt{\_\_lazy\_static\_internal!}           & 64,161  \\
device\_tree\_source & \texttt{parse\_data}            & 17 & trackable    & \texttt{track!}                                & 47,648  \\
npy                  & \texttt{map}                    & 17 & serde        & \texttt{forward\_to\_deserialize\_any\_method} & 43,063  \\ \bottomrule
\end{tabular}
\end{adjustbox}
\end{table}
\vspace{-0.2em}

\paragraph{Macro dispatch}
When comparing the out-degree for both all and inter-package calls, we can
observe a similar trend between them: the median is 1 (mean: 1.7) and the 99th
quantile is 6, suggesting that macro-dispatched calls are largely inter-package
calls. This resonates with our observations for macro-dispatched calls in
\Cref{tab:ds}. Looking at functions calling the most number of macros 
in \Cref{tab:macro:degree}, we can observe that the outdegree generally is
relatively low in comparison to the other two dispatch types. The function
\texttt{path\_segement} in \texttt{item} makes in total 19 macro calls, the
highest in \crates. The median in-degree is 7 (mean: 146) and the 99th quantile
is 1427. When only considering inter-package calls, the median is 12 (mean: 391)
and the 99th quantile is 6433. We can observe comparable numbers to the
in-degree with the other two dispatch types in \Cref{tab:macro:degree}. With
over $200,000$ functions in \crates calling \texttt{log!}, it is the most called
macro followed by \texttt{\_\_impl\_bitflags!} and
\texttt{\_\_lazy\_static\_internal!}. Generally, we can observe that the top
most called macros belong to popular packages in \crates that are known to simplify
logging (\texttt{log}), generate bit flag structures (\texttt{bitflags}), and
wrapping error messages (\texttt{quick-error}).\\

\begin{mdframed}[roundcorner=0pt,nobreak=true,align=center]
The median function in \crates makes one static call, one macro call and has a
\texttt{vtable} with nine function targets. The median
function is also dependent upon by one static call, one macro
call, and implemented by three functions.
\end{mdframed}
\end{sloppypar}

\begin{mdframed}[roundcorner=0pt,nobreak=true,align=center]
\crates is a scale-free network, indicating the presence of a handful of
functions or hubs that are highly connected to other functions in the
repository
\end{mdframed}

\subsection{RQ2: Evolution}

\begin{figure*}[tb]
   \centering
   \includegraphics[width=1\columnwidth]{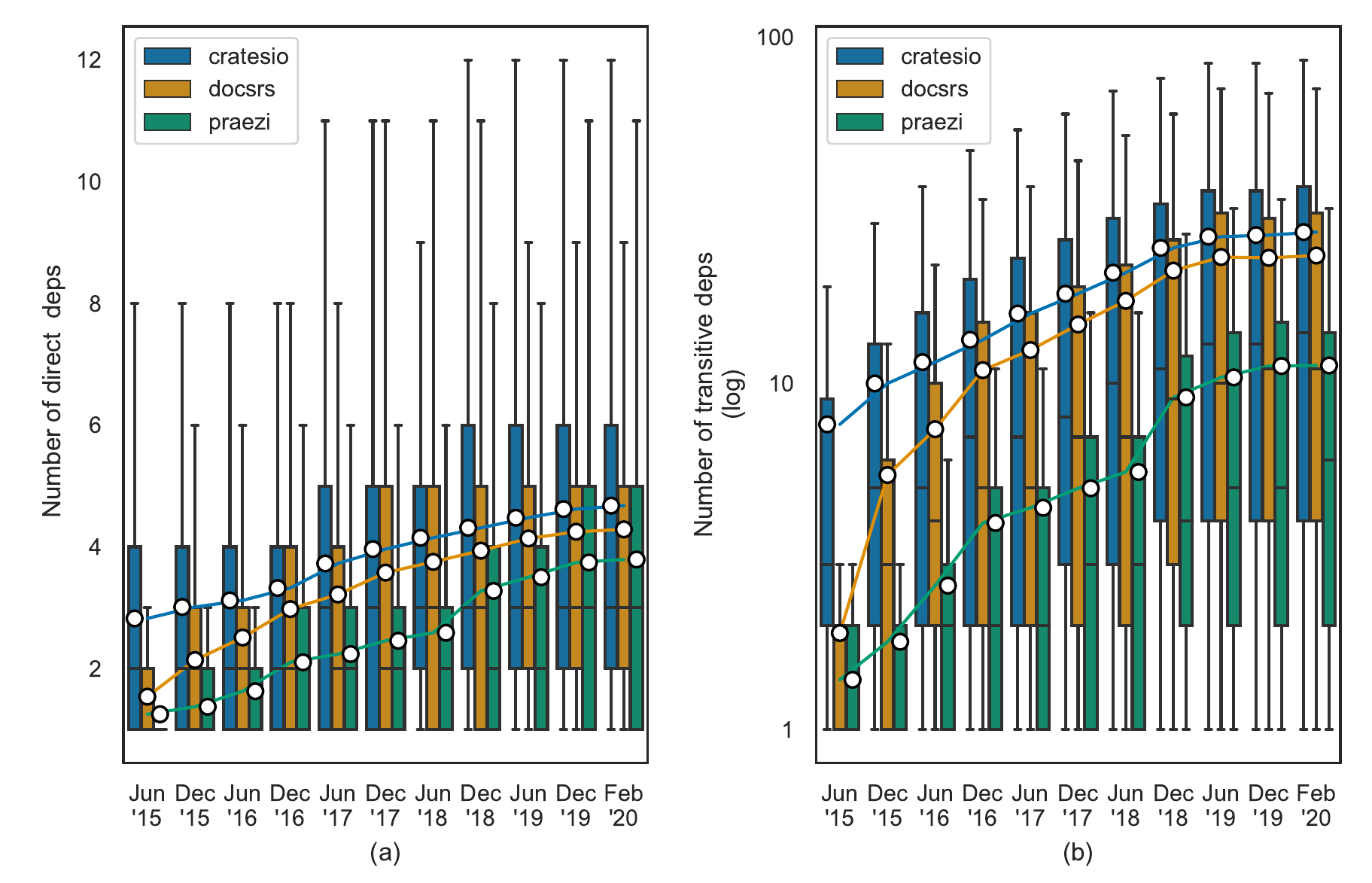}
   \caption{The evolution of package dependencies on two metadata-based networks, \crates and \texttt{Docs.rs},  and one call-based network, \ename.}
   \label{fig:evo:pkg}
\end{figure*}

\subsubsection{RQ2.1: How do package dependencies and dependents evolve?}\label{sec:evo:dep}
\begin{figure*}[tb]
\centering
\includegraphics[width=1\columnwidth]{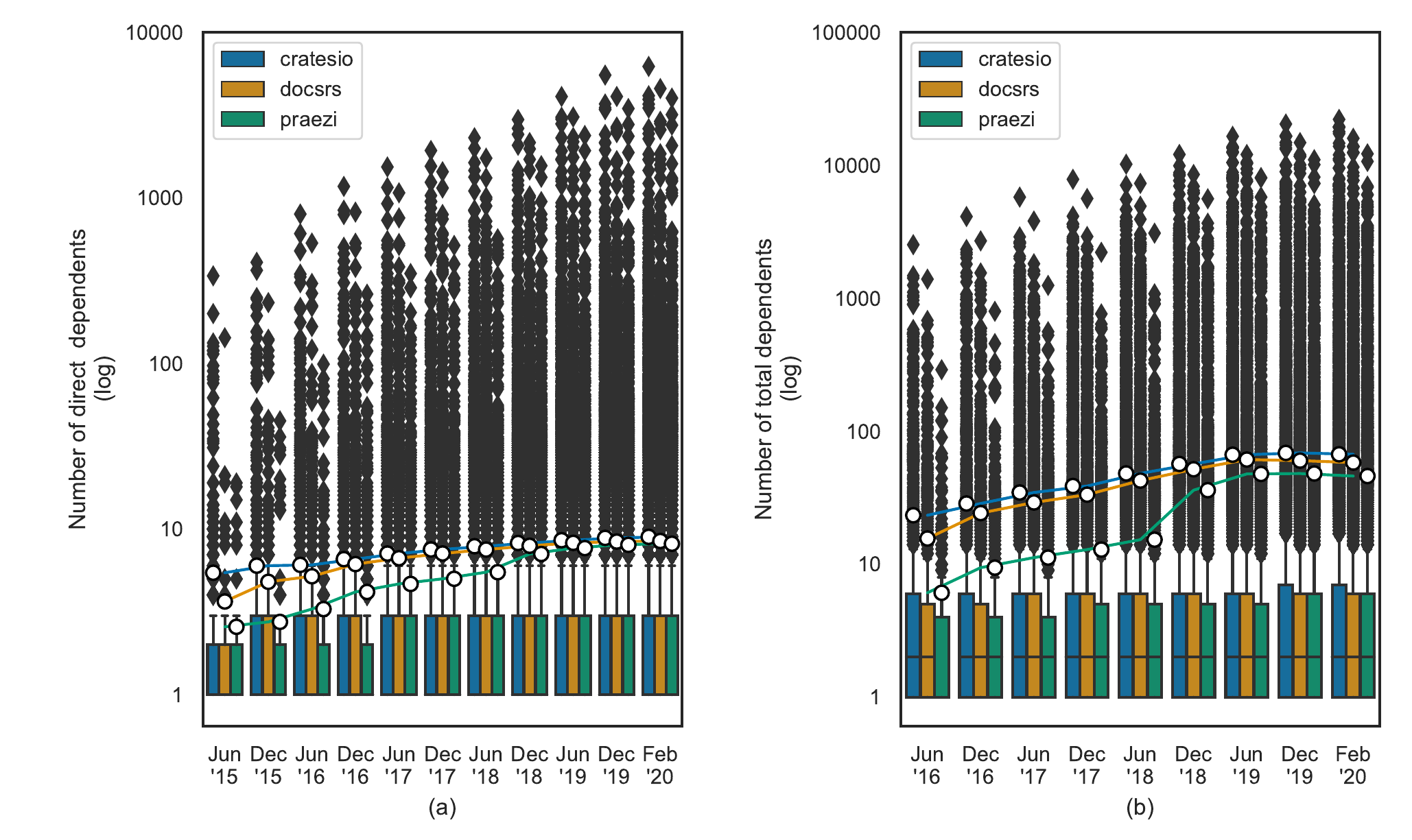}
\caption{The evolution of package dependents on two metadata-based networks, \crates and \texttt{Docs.rs},  and one call-based network, \ename.}
\label{fig:evo:pkg_dependents}
\end{figure*}

\Cref{fig:evo:pkg,fig:evo:pkg_dependents} present
the number of direct and transitive package relationships split by network type
over time. Each sub-plot also features line plots showing the mean with a circle
for each snapshot. By using three different network representations, we can
understand and contrast the differences between the three approximations of
dependency relationships.

\paragraph{Direct dependencies}
Direct dependencies refer to the dependencies that a developer specifies in a
package manifest. For each network group in \Cref{fig:evo:pkg}a, we see a
marginal growth in the median number of direct dependencies over time. The
median number of dependencies for a package grew from two to three between
2015-2020 for the \crates index network as an example. The median is also
similar in the other two networks. Although there are notable differences in the
overall spread in the formative years of \crates, the growth curve is relatively
comparable between the networks. The correlation between the number of direct
dependencies between the three networks (normalized) yields a significantly strong $\rho =
0.89$ between 2017 and 2020 (2015-2017: $\rho = 0.71$), indicating that the
networks approximate each other.

When comparing the mean between the CDN and the \crates index network, we find the
average package call at least one function in 78.8\%\footnote{after normalizing the networks (i.e., inner join of common packages in all three networks)} of its direct dependencies.
As the \crates index network has a higher overall spread than the
\texttt{Docs.rs} network, and the \texttt{Docs.rs} network has a higher overall
spread than the CDN, we can derive that the \crates network represents an
upper-bound and the CDN a lower-bound on the number of direct dependencies. With
75\% of all packages having less than six direct dependencies, the results are
overall similar to the findings of~\cite{decan2019empirical}
and~\cite{kikas2017structure}.

\begin{mdframed}[roundcorner=0pt,nobreak=true,align=center]
Package maintainers use 2 to 3 direct dependencies and are unlikely to
increase their use over time. The three networks have comparable results.
\end{mdframed}

\paragraph{Transitive dependencies}
Transitive dependencies represent the indirect dependencies of a package after
resolving its specified dependencies. In comparison to the direct dependencies,
in \Cref{fig:evo:pkg}b, we can observe an initial superlinear growth, followed
by a period of stabilization (since 2018) for the three
networks. The median number of transitive dependencies in 2015 is 5
for the \crates index network and 1 for the other two networks. The median number of
transitive dependencies grew with a delta of 5 additional packages for the CDN,
9 for the \texttt{Docs.rs} network, and 12 for the \crates index network in five
years. While we can find a similar continuing growth trend to
\Cref{fig:evo:pkg}a, we observe higher degrees of
dispersions between the CDN and the other two networks. The third-quartile in
nearly all CDN snapshots is the same or below the median of the other two
networks. Thus, half the packages in the \crates index network and
\texttt{Docs.rs} network report a higher number of transitive dependencies than
75\% of packages in the CDN. When normalizing the networks and comparing the mean between the CDN and the
\crates index network in 2020, we find the average package call at-last one function in 40\%\footnote{See footnote 22} of all
resolved transitive dependencies. The discrepancy indicates substantial
differences between call-based and metadata-based networks in network analyses;
CDNs will overall report a notably lower number of transitive dependencies than
the metadata-based ones. 

Finally, the correlation between the number of transitive dependencies between
the three networks (normalized) is generally strong, with an average $\rho = 0.84$ between
2017 and 2020 (2015-2017: $\rho = 0.70$). In other words, the more resolved
transitive dependencies a package has, the more transitive dependencies it will call (and vice versa). However, we identify a moderate
average correlation $\rho = 0.62$ between the number of direct dependencies
(i.e., either metadata-based or call-based) and the number of call-based
transitive dependencies in 2017-2020. In 2015-2017, we observe a general weaker
correlation, with $\rho = 0.47$. Thus, two packages with the same number of
direct dependencies are likely to have different number of transitive
dependencies.

\begin{mdframed}[roundcorner=0pt,nobreak=true,align=center]
The average dependency tree of resolved packages has nearly grown thrice (5 to
17 transitive dependencies) in 5 years. Substantial differences exist between the
networks; packages are not calling 60\% of their resolved transitive dependencies. 
\end{mdframed}

\paragraph{Direct dependents}\label{sec:evo:dependents}

In addition to dependencies, dependents measure the number
of consumers a package has. In the context of this study, we consider a consumer
as an internal consumer (i.e., a package making use of another package within
\crates). \Cref{fig:evo:pkg_dependents}a presents the number of dependents over
time. Irrespective of the network, we can see that the median number of
consumers per package remains unchanged at one over time. Similarly, we can also
find the interquartile ranges of the networks to be identical from June 2017 and
onwards. In that period, the top 25\% packages have at least three or more
consumers. The correlation between the number of direct dependents for the
three networks (normalized) yields a strong $\rho = 0.81$ between 2017 and 2020 (2015-2017:
$\rho = 0.75$), indicating (similar to direct dependencies) that the networks
closely approximate each other.

When comparing the mean over time, we see a steady growth of the
number of direct dependents for all three networks. The growth pattern is a
result of a few commonly used packages (e.g., \texttt{serde} and \texttt{log})
having the largest share of consumers in \crates (see also
\Cref{fig:degree:downstream}). The outliers in the boxplot represents the
top-most used packages for each network. Here, we can observe
notable differences in the range and number of outliers between the networks.
The number of top dependent packages in June 2018 is 651 for the CDN, 1245 for
the \texttt{Docs.rs} network, and 1680 for the \crates index network. There are
2.5x more top-dependent packages for \crates than in the CDN. When comparing the
top-most dependent packages in each network, the most consumed package has 566
dependents in CDN, 1735 in the \texttt{Docs.rs} network, and 2305 in the \crates
index network. Although the gap between the outliers in the networks reduces
over time (i.e., from 2.5x to 1.8x in 2020), there are notable differences
between the networks when analyzing the top-most dependent packages in \crates.

Overall, the results are similar to the
findings of both \cite{decan2019empirical} and \cite{kikas2017structure},
suggesting that an average \crates package has a relatively constant and low
degree of consumers in general. While the networks seem comparable and interchangeable at
large, there is a notable discrepancy between the outliers (i.e., topmost used
packages in \crates) in metadata-based networks and call-based networks in
earlier snapshots, potentially yielding differences in network analyses of top
dependent packages.\\

\begin{mdframed}[roundcorner=0pt,nobreak=true,align=center]
The average number of consumers of a package remains at one over time. Similar
to direct dependencies, the networks approximate each other (except for
top-dependent packages).
\end{mdframed}

\paragraph{Total dependents}
\Cref{fig:evo:pkg_dependents}b shows the total number of dependents per package.
The total number of dependents include both direct and transitive dependents. We
omit both June and December 2015 as these snapshots only have 19 and 47
transitive dependents in the CDN, respectively. Except for June 2016, the median
number of total dependents remains constant at two for the three networks. Thus,
in addition to the one median direct consumer in \Cref{fig:evo:pkg_dependents}a,
packages also have one median transitive consumer. When looking at the top 25\%
consumed packages, the number of total dependents ranges from 8 or more
consumers for the \crates index network and 7 or more consumers for the
remaining networks. There is also a slight increase in the overall range at two
occurrences for the CDN (Feb'17, Dec'19) and one occurrence for the
\texttt{Docs.rs} network (Dec'17) and the \crates index network (Dec'19). When
comparing the mean and outliers between the networks, we find a similar growth
pattern and gap to \Cref{fig:evo:pkg_dependents}a. 

Similar to transitive dependencies, we also find a general strong correlation
between the number of transitive dependents between the three networks (normalized)
($\rho = 0.77$), and also a moderate correlation between the number
of direct dependents and transitive dependents ($\rho = 0.54$).

Overall, we see that the total number of dependents
remains stable over time with a few cases of gradual increase. Moreover, we
see that the distributions of dependents are generally much lower in comparison
to the transitive dependency relationships in \Cref{fig:evo:pkg}b. Thus, the
results indicate that an average package in \crates has a handful stable number
of consumers.

\begin{mdframed}[roundcorner=0pt,nobreak=true,align=center]
The average package also has one transitive consumer that remains unchanged
over time. Similar to direct top-most dependent packages; indirect consumers are
using them to a much higher degree than previously.
\end{mdframed}

\begin{figure*}[tb]
   \centering
   \includegraphics[width=1\columnwidth]{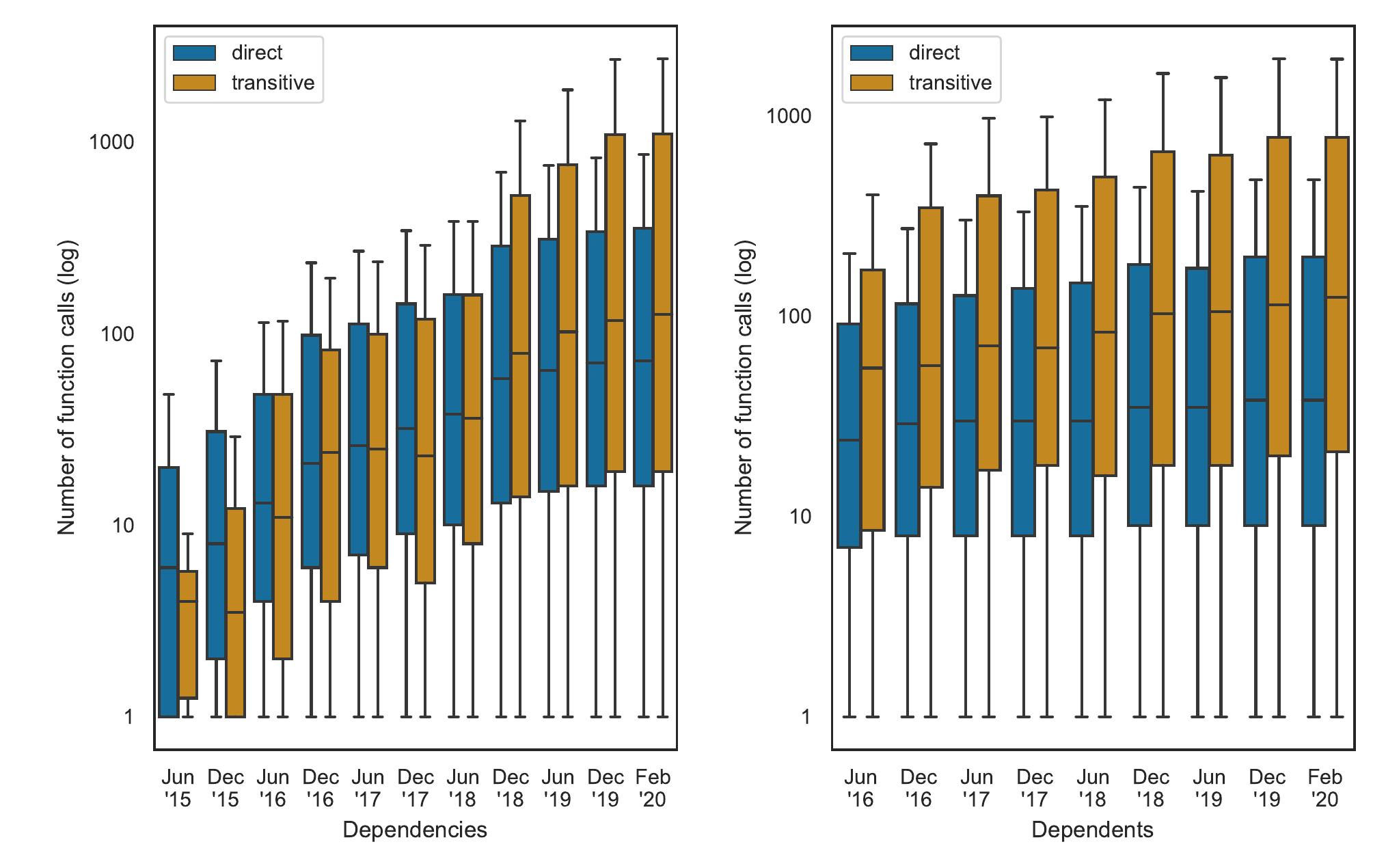}
   \caption{The evolution of the number of functions calls to dependencies and dependents}
   \label{fig:evo:pkg-fns}
\end{figure*}

\subsubsection{RQ2.2: How does the use of external APIs in packages evolve?}\label{sec:evo:fns}
\Cref{fig:evo:pkg-fns} describes the evolution of the
number of direct and transitive inter-package (i.e., API) calls per package for
dependencies on the left-hand side and dependents on the right-hand side. When
looking at the number of calls to dependencies over time, we make two major
observations. First, the number of direct and transitive calls to dependencies
has an initial superlinear growth, followed by a period where the growth slows down
from December 2018 and onwards. From December 2016 to December 2019, the number
of direct calls grew from 21 (transitive: 24 ) to 70 (transitive: 116),
a three-fold increase in three years. On average, we also see a growth of 6.6 new
function calls to direct dependencies and 12.2 new indirect calls to transitive
dependencies every six months. Second, we can see that the median number of
transitive calls overtakes the median number of direct calls in
December 2018. Our findings unveil that the amount of calls to indirect APIs
are comparable in numbers to calls of direct APIs. Recent snapshots
further indicate that packages invoke more indirect APIs than direct APIs. The
transitive median API calls in December 2019 is 1.6x larger than the
median direct API calls.

\begin{mdframed}[roundcorner=0pt,nobreak=true,align=center]
The average API usage of transitive dependencies is both greater and comparative to direct dependencies in recent years.
\end{mdframed}

Similar to the total dependents in \Cref{fig:evo:pkg}d, we also omit the two
snapshots in 2015 due to an insignificant number of transitive dependents.
Generally, we can observe a continuous growth of the number of direct and
transitive consumers of package APIs over time. The median number
of consumer grew from 25 callers in 2015 to 38 callers in 2020, an average growth of
1.6 new functions per year. The median of indirect consumers is
larger than that of direct consumers, growing from 55 callers in 2015 to 124 callers in 2020, an
average of 8.75 new functions every six months. When comparing the growth pattern
between direct and transitive dependents, the gap between the
median of direct dependents and transitive dependents expands over time.
Moreover, we also find that the interquartile ranges and overall range is
greater for transitive dependents than for direct dependents in all snapshots. A
package with transitive dependents is likely to have more indirect callers than direct callers of their
APIs. Notably, the median number of transitive dependent callers (median: 114) is
three times larger than the median of direct dependent callers (median: 38) in 2020. When
also taking into account the findings of transitive dependency callers, our
results strongly indicate that indirect users of library APIs is both highly
prevalent in \crates, and comparable to direct users of library APIs. Despite
the largely unchanged number of direct and total dependents (See
\Cref{fig:evo:pkg_dependents}) over time, we see indications that developers are
increasingly using more APIs over time. 

\begin{mdframed}[roundcorner=0pt,nobreak=true,align=center]
Packages with transitive consumers have three times more API callers stemming from indirect
consumers than direct consumers.
\end{mdframed}

Below, we summarize the two perspectives of package relationships using both the
metadata-based results with function-based results:

\begin{mdframed}[roundcorner=0pt,nobreak=true,align=center]
\textit{Dependencies:} Packages depend on an increasing number of transitive
dependencies over time. Package maintainers, however, are not declaring more
dependencies. Although there is an increase of new direct and indirect API calls
to dependencies over time, roughly 60\% of all resolved transitive dependencies
are not called. 
\end{mdframed}

\begin{mdframed}[roundcorner=0pt,nobreak=true,align=center]
\textit{Dependents:} The number of total dependents, one direct and one
transitive consumer, remains constant over time. However, consumers have a
growing number of callers over time. For packages with transitive consumers,
there is a higher number of calls stemming from indirect callers than direct
callers.
\end{mdframed}

\begin{figure}[tb]
\centering
\includegraphics[width=0.75\columnwidth,keepaspectratio]{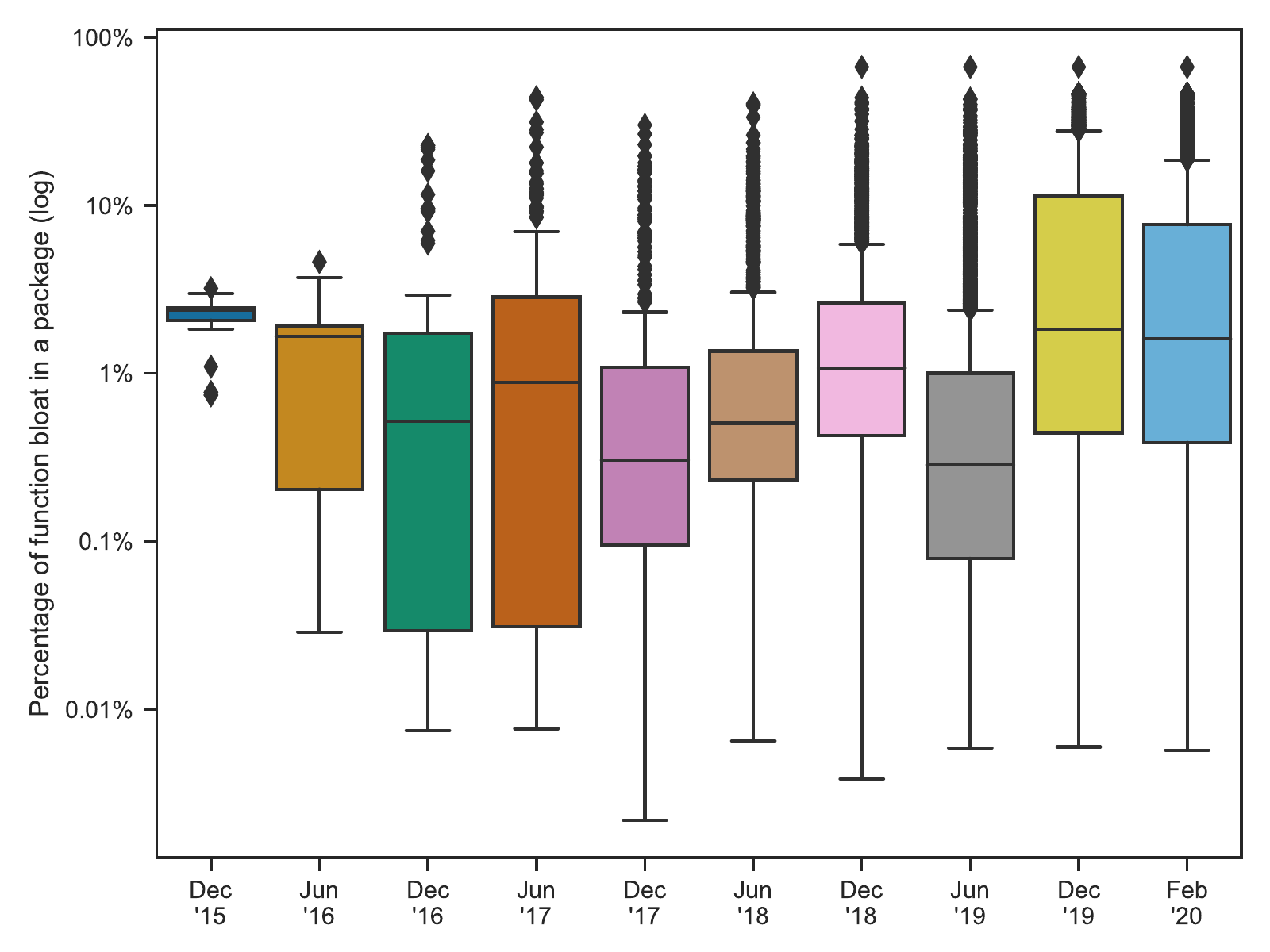}
\caption{Percentage of co-existing functions (i.e., bloat) in \crates packages}
\label{fig:bloat}
\end{figure}

\subsubsection{RQ2.3: How prevalent is function bloat in package dependencies?}\label{sec:evo:coexist}
Packages depending on a growing number of external packages are also likely to
introduce dependency conflicts. Conflicts arise when a dependency resolver is
unable to eliminate the co-existence of a package in a dependency tree due to
version incompatibility. For example, a resolver may arrive that there is no
overlapping version when two packages in a dependency tree depend on package \texttt{A}
where the former specifies a version constraint \textit{$1.*$} and the latter
\textit{$2.*$}. Rust's \cargo package manager avoids such conflicts by allowing
multiple versions of the same package to co-exist in a dependency tree using
\textit{name mangling} techniques~\citep{blogrustcargo}. A potential drawback of
this strategy is the risk of bloating binaries due to multiple copies of
identical yet obfuscated functions. 

As a proxy for function bloat in binaries, we
calculate the percentage of co-existing functions for all public functions in
\crates. We denote a co-existing function as multiple copies of identical
function identifiers loaded from different versions of the same package. It is
important to note that the measure is an estimation and does not guarantee the
semantic equivalence of functions. Before measuring the percentage of
co-existing functions, we first inspect the presence of co-existing functions
in all \crates packages. On average, we find packages having at least one
co-existing function to be 5.4\% of \crates in Dec 2015-Dec 2017 and 28\% of
\crates in Jun 2018-Feb 2020. There are no packages with co-existing functions
in June 2015. Largely non-existent in the formative years of \crates, we find
that function co-existing among dependencies is relatively prevalent in recent
years.

Among packages having co-existing functions, \Cref{fig:bloat} breaks down the
percentage of co-existing functions in dependencies of packages over time. We
can observe that the median fluctuates between 0.3\% and 1.6\% over time,
indicating a constant yet insignificant amount of function co-existence in
packages. 75\% of all packages range between 1 to 10\% co-existing functions in
their dependencies, suggesting that a majority of packages have a small amount of
possible bloat in their binaries. Thus, bloating of binaries from co-existing
dependency functions are highly unlikely for packages with at least one
co-existing function in \crates.

Finally, we find a small minority (i.e., outliers) of packages with a high
degree of possible function bloat between December 2018 and February 2020. The
package reporting the highest bloat of this time frame is \texttt{downward} with
67\% bloat. However, it is an invalid outlier as it has a circular dependence on
itself.\footnote{\url{https://crates.io/crates/downwards}} Thereby, the two
packages with highest bloat is \texttt{const-c-str-impl} and \texttt{mpris} with
43\% and 46\% bloat, respectively. Upon manual inspection of their respective
dependency tree, we identify that the packages have a dependence on multiple
versions of \texttt{proc\_macro}, \texttt{quote}, \texttt{syn}, and
\texttt{unicode\_xid}, common libraries for creating procedural macros. For
example, \texttt{mpris} indirectly uses four different versions of \texttt{syn} and
\texttt{quote}.\footnote{\ahref{http://web.archive.org/web/20201201224352/https://docs.rs/crate/mpris/2.0.0-rc2/source/Cargo.lock}{https://docs.rs/crate/mpris/2.0.0-rc2/source/Cargo.lock}}.
We also make similar observations in three other outliers: \texttt{js-object}
(33\%), \texttt{js-intern-proc-macro} (41\%), and \texttt{mockers\_derive}
(43\%). Further investigation could perhaps  reveal whether the combination of
certain procedural macros libraries are highly likely to always result in bloated
dependency tree configurations.

\begin{figure*}[tb]
\centering
\includegraphics[width=0.7\columnwidth,keepaspectratio]{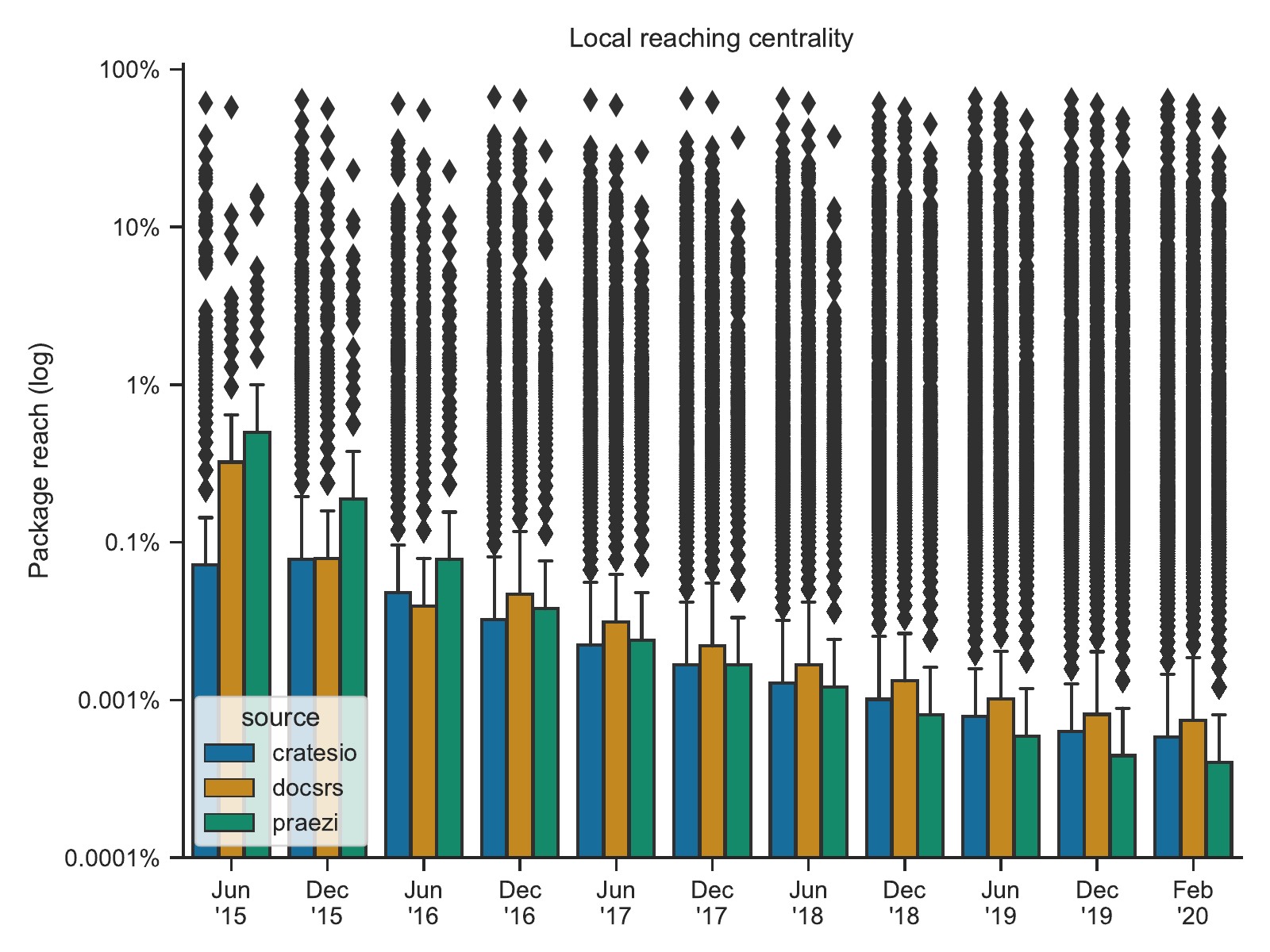}
\caption{Distribution of Package Reachability}
\label{fig:reach}
\end{figure*}
      
\begin{table}[tb]
\caption{Most central APIs in the largest component in  Dec 2015, Dec 2017, and Feb 2020}
\label{tab:api:stats}
\centering
\begin{adjustbox}{width={1\columnwidth},keepaspectratio}
\begin{tabular}{llr|llr|llr@{}}
\toprule
\multicolumn{3}{c|}{Dec 2015 (size: 534)}
& \multicolumn{3}{c|}{ Dec 2017 (size: 6,004)}
& \multicolumn{3}{c}{ Feb 2020 (size: 24,857)}
\\
Package           & Function               & \multicolumn{1}{l|}{Reach} & Package              & Function         & \multicolumn{1}{l|}{Reach} & Package             & Function         & \multicolumn{1}{l}{Reach} \\ \hline
pkg-config::0.3.6 & \texttt{find\_library}        & 10\%                       & log::0.3.8           & \texttt{\_\_log}          & 16\%                        & log::0.4.10         & \texttt{max\_level}       & 30\%                        \\
gcc::0.3.20       & \texttt{Build::new}           & 6\%                        & libc::0.2.34         & \texttt{memchr}           & 12\%                        & serde::1.0.104      & \texttt{next\_element}    & 24\%                       \\
libc::0.2.1       & \texttt{memchr}               & 6\%                        & lazy\_static::0.2.11 & \texttt{get}              & 11\%                        & bitflags::1.2.1     & \texttt{\_\_fn\_bitflags} & 23\%                       \\
log::0.3.4        & \texttt{\_\_static\_max\_level} & 6\%                        & bitflags::1.0.1      & \texttt{\_\_fn\_bitflags} & 8\%                        & lazy\_static::1.4.0 & \texttt{get}              & 21\%                       \\
bitflags::0.3.3   & \texttt{bitflags}               & 3\%                        & unicode-width::0.1.4 & \texttt{width}            & 8\%                        & libc::0.2.66        & \texttt{sysconf}          & 18\%                      \\
gcc::0.3.20       & \texttt{Build::compile}        & 3\%                         & serde::1.0.24         &\texttt{deserialize}     & 6.3\%                      & libc::0.2.66         & \texttt{isatty}       & 18\%                        \\
log::0.3.4        & \texttt{log::macros log}        & 6\%                        & lazy\_static::0.2.11       & \texttt{lazy\_static}  &  5.95\%                 & memchr::2.3.2      & \texttt{memchr}    & 18\%                       \\
gcc::0.3.20       & \texttt{compile\_library}       & 4\%                        & byteorder::1.2.1  & \texttt{write\_u32}            &  5.1\%                   & itoa::0.4.5      & \texttt{Buffer::new} & 17.5\%                       \\
time::0.1.34      & \texttt{precise\_time\_ns}       & 2.5\%                     & libc::0.2.34      & \texttt{localtime\_r} &  5\%                              & ryu::1.0.2   & \texttt{Buffer::format\_finite}               & 17\%                       \\
libc::0.2.1       & \texttt{sysconf}               & 2.5\%                       & time::0.1.38    & \texttt{num\_seconds}             & 4.1\%                   & serde\_json::1.0.48  & \texttt{from\_str}         & 10\%                      \\ \bottomrule
\end{tabular}
\end{adjustbox}
\end{table}

\begin{mdframed}[roundcorner=0pt,nobreak=true,align=center]
28\% of all packages in \crates have a co-existing function in their
dependencies. Among those packages, between 1-10\% of imported functions from
dependencies are bloated.
\end{mdframed}

\subsubsection{RQ2.4: How fragile is \crates to function-level changes?}\label{sec:evo:coexist}
Our goal is to identify packages that indirectly reach most of
\crates and understand the differences and similarities in using different
networks for impact analyses of package repositories. We use the local reaching
centrality~\citep{mones2012hierarchy} to measure the reach of individual
packages in the CDN, compile-validated metadata (i.e., \texttt{Docs.rs}), and
regular metadata (\crates) networks. With reach, we measure the fraction of
\crates packages that depend on a particular package (i.e., its transitive dependents). 

\Cref{fig:reach} presents the evolution of the reach of each package per
network. When comparing the third-quartile between the snapshots, we can observe
a gradual decrease in reachability over time. The decrease is a result of new packages being added to the network and at the same time not being widely used by other packages. The top 25\% of the distribution of the \crates index
network has a ten-fold decrease of 0.07\% in June 2015 to 0.008\% in June 2019.
Both the CDN and \texttt{Docs.rs} distribution also follow a similar pattern. In
comparison to recent years, the higher reach of packages in the formative years
reflects the small network size. In the remaining 75\% of packages, they have no
or limited reach of \crates irrespective of network choice, indicating that a
majority of packages do not exhibit any influence in \crates. However, we can
observe that the range and number of outliers expand over time, indicating that
there is an increasing number of packages that exhibit a degree of influence in
\crates. The number of outliers with greater than 10\% reachability grew from 19
(\texttt{Docs.rs}/CDN: 3) to 92 (\texttt{Docs.rs}: 80, CDN: 66) packages as an
example. 

For each snapshot, we can see that the top-most outlier and the number of
outliers is lower than that of the metadata-based network in each network. The
most reachable package in June 2019 reaches 65\% in the \crates index network,
61\% in the \texttt{Docs.rs} network, and 47\% in the CDN network. 

Upon inspection of the top 10 highest reaching outliers in each network, we see that
a similar set of packages such as \texttt{libc}, \texttt{log},
\texttt{lazy\_static}, and \texttt{bitflags} remains prominent over time across
the networks. These packages are also among the most directly called packages in
\Cref{sec:desc:fns}. \texttt{libc}, one of the most downloaded packages in
\crates, is the package exhibiting the highest all-time influence in \crates.
There are also packages in decline: \texttt{rustc-serialize}, a serializer
package, decreased in reach from its peak of 17\% in 2016 to 2\% in 2020. A
potential explanation for its decline could be the adoption of \texttt{serde}, a
rivaling serializer package, that grew its reach from 6\% in 2016 to 42\% in
2020. 

We derive the ten most influential APIs by measuring the local reach centrality
on functions of the CDN for 2015, 2017, and 2020 in
\Cref{tab:api:stats}.\footnote{due to presentation reasons, we showcase for only
three years} Although \texttt{libc}
exhibit the highest reach at the package-level, functions in \texttt{log} or
\texttt{serde} exhibit higher influence than individual functions in
\texttt{libc}. Moreover, we can see that \texttt{libc}, \texttt{log}, and
\texttt{bitflags} have remained important since the inception of \crates.
However, we can observe that the most called function changes over time. For
example, \texttt{log} reports three distinctly different API functions. A
possible explanation could be that new features or best practices over time
change the use of APIs. Finally, we can also see a new fast-growing entrant in
2020: \texttt{serde} is second to \texttt{log}.

\begin{mdframed}[roundcorner=0pt,nobreak=true,align=center]
A large majority of packages in \crates have no or limited reachability; a
handful of packages are reachable from 47\% of \crates, and single functions are
reachable from 30\% of \crates.
\end{mdframed}

\subsection{RQ3: Reliability}
We identify two occurrences with significant differences between the studied
networks, namely transitive dependencies and outliers in the top-most dependent
packages in \textbf{RQ2.1}. These differences have practical implications on
dependency analysis use cases. For example, security-based dependency analysis such as
\texttt{cargo-audit} would generally favor soundness over precision. Failing to
account for an actual dependency relationship could lead to vulnerabilities
being undetected. On the other hand, automated dependency updating such as
\github's \texttt{Dependabot} would favor precision over soundness.
False-positive updates steal valuable time from
developers~\citep{mirhosseini2017can,beller2016stateofasats}. Thus, our goal in
\textbf{RQ3} is to obtain an understanding of how accurate and reliable
metadata-based and call-based networks are in estimating actual relationships
between packages. 

\paragraph{Selection}
As packages can have many transitive dependencies and have complex use cases,
manually mapping out how packages use each other in a dependency tree is a
tedious and error-prone task. Attempting to scale the analysis to the entire
\crates is also impractical. Thus, we sample dependency
relationships in packages where both the metadata-based networks and the
call-based networks report differently (e.g., between a package and a
dependency, the metadata-network reports an edge between them, and the call-based
network does not). We can then focus our manual investigation on whether
call-based networks are missing function calls due to limitations with static
analysis or whether metadata-based networks
over-approximate unused dependencies. Moreover, analyzing a narrow set of direct
and transitive dependencies further reduces the overhead of manually tracking
uses of code elements across packages and their dependencies. 

In the span of five workdays, we randomly sampled and reviewed 34 cases, 7 cases
involving direct relationships, and 27 cases involving transitive relationships.

\paragraph{Review Protocol}
We initiate the review by first finding import statements of the
direct library for the package under analysis and then track successive uses of
imported items in variable assignments and definitions such as functions (e.g.,
return type) and \texttt{trait} implementations. After mapping out all use
scenarios that trace back to the original set of import statements, we later can
conclude whether a package reuses code from a dependency. The procedures for
direct and transitive dependencies are slightly different. For direct
dependencies, we investigate the entire package for any sign of reuse. For
transitive dependencies, we inspect the context of how a package reuses its
direct dependency, and whether the specific reuse of the direct dependency leads
to reuse of the transitive dependency. Given the following example: Package
\texttt{Foo} depends on \texttt{Bar}, and \texttt{Bar} depends on \texttt{Baz}.
\texttt{Foo} also reuses \texttt{Bar}, and \texttt{Bar} also reuses
\texttt{Baz}. A function \texttt{bar()} in \texttt{Bar} calls \texttt{baz()} in
\texttt{Baz} and \texttt{foo()} in \texttt{Bar} does not rely on external code.
If \texttt{Foo} only calls \texttt{foo()}, then \texttt{Foo} only reuses
\texttt{Bar} and not \texttt{Baz} despite \texttt{Bar} reusing \texttt{Baz}. If
\texttt{Foo} would call \texttt{bar()}, then there is an indirect reuse of the
transitive dependency \texttt{Baz}. A step-by-step review protocol is
available in the replication package.

\paragraph{Manual Analysis}

\begin{table}[tb]
\caption{Manual inspection and classification of 34 dependency relationships between \ename and the \crates index network.}
\label{tab:manual}
\centering
\begin{tabular}{@{}lr@{}}
\toprule
\bfseries{Categorization} & \bfseries{\#Samples} \\
\midrule
i) Over-approximation in metadata-based networks & \textbf{27} \\
~~~~~~~\ldots~no import statements & 3 \\
~~~~~~~\ldots~import statement and no usage &  4 \\
~~~~~~~\ldots~resides in a \texttt{\#[cfg(...)] block} & 1 \\
~~~~~~~\ldots~derive macro libraries  &  2 \\
~~~~~~~\ldots~test dependency  &  1 \\
~~~~~~~\ldots~non-reachable transitive dependency & 16  \\
\midrule
ii) Under-approximation in \ename & \textbf{7}\\
~~~~~~~\ldots~importing a constant & 1 \\
~~~~~~~\ldots~importing data type and usage & 1  \\
~~~~~~~\ldots~importing data type in definitions & 4 \\
~~~~~~~\ldots~handling C-function call & 1 \\
\midrule
$\Sigma$ & \textbf{34} \\
\bottomrule
\end{tabular}
\end{table}

\Cref{tab:manual} tabulates the reasons for misclassification split by network
and number of use cases. Overall, the metadata-based network over-approximates
the dependency usage in 80\% of the analyzed cases. Among direct dependencies
where the metadata-based networks over-approximate, we identify seven instances
where a package did not import any item from the dependency relationship under
analysis.  Moreover, metadata-based
networks cannot distinguish dependency usage in non-runtime or conditionally
compiled sections of the source code.  We found two cases; one case where a
developer uses a runtime dependency solely in test code and one
conditional compilation case where a dependency code runs only on Windows environments.

While \cargo has labels for build, test, optional, and platform-specific
dependencies in the manifest file,~\textit{derive macro} dependencies are not
distinguishable from runtime dependencies. A \textit{derive macro} library performs
code-generation at compile time. However, such libraries do not provide runtime
functionality and are closer to the role of being a build dependency. We identify two
such libraries, \texttt{cfg-if} and \texttt{thiserror}. Including such
dependencies influences the count of runtime dependencies; for example,
depending on the widely popular
\texttt{serde\_derive}\footnote{\url{https://docs.rs/crate/serde_derive/1.0.106/source/Cargo.lock}}
library would incorrectly add six dependencies to the total count of runtime
dependencies. Without no specific metadata label or heuristic, a call-based
dependency network avoids including such libraries. 

\begin{sloppypar}
The most prominent case with over-approximation by metadata-based networks are
non-reachable transitive dependencies. The context of how a
package uses its direct dependencies plays a central role in whether a package
indirectly uses its transitive dependencies. As an example, the package
\texttt{selfish} uses \texttt{nom v3.2.1} that then depends on \texttt{regex
0.2.11}. \texttt{nom} is a parser library and exports a set of regex parsers
that uses the \texttt{regex} library. Although \texttt{selfish} enables the
regex feature in \texttt{nom}, it does not import any of the regex parsers in
\texttt{nom}, effectively making the \texttt{regex} library unused. 
\end{sloppypar}

In the four cases where a developer imports type definitions from dependencies
for use in function declarations. One such example is the case of importing
\texttt{c\_int} in \texttt{libc}  for  function declarations in	\texttt{whereami
v1.1.1}.  Although a call graph does not track data references, we could still
mitigate this by tracking the type declarations in argument and return types of
functions in the call graph. \ename embeds full type qualifiers including
package information in functions belonging to call-based dependency networks
(See \Cref{cg:annot}).

Finally, we identify one instance where the call graph
generator could not resolve a call from \texttt{subprocess v0.1.0} to the
\texttt{libc} function \texttt{pipe()}. Although there is a \texttt{pipe} call
without clear identifiers in the call graph, it is not via the \texttt{libc}
library. Thus, there are possible limitations with handling cross-language calls.

\begin{mdframed}[roundcorner=0pt,nobreak=true,align=center]
A call-based dependency network is more precise than a metadata-based network.
Data-only dependencies could affect its soundness. 
\end{mdframed}

\section{Discussion}
We center our discussion on two key aspects; differences and similarities
between using three different networks for network analyses and studying
function relationships on a network level. 

\subsection{Strengths and Weaknesses between Metadata and Call-based Networks}

As package repositories do not test whether a package can build or not,
developers can by mistake or unknowingly publish broken versions to \crates. By
verifying the build of package releases, the \texttt{Docs.rs} network excludes
package releases that do not have a successful build record. When comparing the
results of the network analyses in \Cref{fig:evo:pkg} with \crates index
network, overall, we find them to have comparable results except in the formative
years of \crates. The diverging results in the initial years show that a large
number of releases are not reproducible and consumable, stressing the importance
of performing additional validation besides the correctness of packages
manifests. Thereby, we urge researchers to minimally validate package manifests
with external information such as publically available build and test data for
network studies of package repositories.

When comparing the network analysis results in \Cref{fig:evo:pkg}, we find
notable similarities and differences between metadata-based and call-based
networks for \crates. Except for the formative years of \crates, the
distributions of recent snapshots for direct dependencies, direct dependents,
and total dependents are mostly similar between the networks. Thus, a network
inferred from \crates metadata closely approximates the presence of function
reuse relationships between packages without needing to construct and verify
with call graphs. Recent snapshots of \crates further indicate that recent
package releases are highly likely to be reproducible and compile as well. On
the other hand, there are also significant differences between the networks,
specifically for transitive dependencies and outliers in dependent
distributions. By taking into account that a developer does not make use of all
APIs available in a package, we identify a two-fold difference between
call-based and metadata-based networks. These differences also manifest among
the most popular dependent packages (i.e., outliers)---despite the networks
reporting similar results for the average dependent package. 

Based on these similarities and differences, we conduct a manual analysis to
understand which network has a more accurate representation of package
repositories. Our investigation indicates that call-based dependency networks
are more precise than metadata-based networks; the prominent finding is that the
number of transitive dependencies a package uses is highly contextual and
moderately correlates with the number of declared dependencies. From a
statistical viewpoint, we identify a strong correlation between the number of
dependencies derived from a metadata-based network and the number of called
dependencies. In other words, the more resolved transitive dependencies a
package has, the more transitive dependencies it will call. On the other hand,
we only observe a moderate correlation between declared (direct) dependencies
and called transitive dependencies, indicating that the number of called
transitive dependencies potentially varies for the same number of direct
dependencies. Based on our studied use cases, we find examples of packages only
importing non-core functionality from libraries or
specific modules of packages that use individual libraries by themselves.
Despite limitations with data-only dependencies, we argue that calculating the
number of transitive dependencies should not be generalized to the sum of all
resolved dependencies. In line with previous work on the fine-grained analysis
of known security vulnerabilities, we also argue both researchers and
practitioners interested in understanding how developers or programs use
dependencies should account for its context---not the number of compiled
dependencies. 

As a summary, we make the following recommendations based on the trade-offs and
costs for constructing a call-based dependency network: 

\begin{sloppypar}
\begin{itemize}
   \item \textbf{Direct dependencies:} Given the relative proximity of results
   between a metadata-based and call-based network, a metadata-based network is
   sufficient for use cases involving direct dependencies if precision is not
   crucial. The cost of building a call-based dependency network would be overly
   expensive.
   \item \textbf{Transitive dependencies:} Where transitive dependencies are
   central in any analysis, we recommend call-based dependency networks
   over metadata-based networks.
   \item \textbf{Data-dependencies: } Where data references are crucial to track
   or studying data-centric packages in \crates, we recommend metadata-based
   dependency networks or use additional (cheap) static analysis to identify
   data dependencies. Although metadata-based networks are imprecise, they will not miss such
   relationships.
\end{itemize}
\end{sloppypar}
\subsection{Transitive API Usage}
\begin{sloppypar}

For studying the evolution, impact, and the decision-making of
deprecation~\citep{robbes2012developers,Sawantdeprecation} and
refactorings~\citep{kula2018empirical} of library APIs, datasets such as
\texttt{fine-GRAPE}~\citep{sawant2017fine} provide valuable insights into how a
large number of clients in the wild make use of a few popular libraries. These
datasets extract API usage by mining direct invocation of library APIs (i.e., a
client calling a public API function). By analyzing the use of APIs in
transitive dependencies of clients (i.e., indirect API use) in addition to
direct dependencies, we find that there are more calls to transitive
dependencies than direct dependencies in recent years. Thus, the transitive
relationship where either an intermediate client or library relays a call
between a client and a library could potentially present new confounding
variables and implications to the evolution and decision-making of APIs.
Although developers do not have control of transitive package dependencies, they
have the same execution rights and follow the same laws of software
evolution~\citep{lehman1980programs} as direct dependencies. Thus, API decisions
in transitive dependencies can equally impact clients as direct dependencies.

As package managers allow the same dependency (albeit different
versions of them) to co-exist in a client, our results in \textbf{RQ2.4} show
growing signs that more and more copies of the same function identifier from
multiple versions exist in a client. In cases where such a function is dependent
on the environment (e.g., a specific implementation of an OpenSSL library),
there is a potential risk for introducing unexpected incompatibilities. Such
problems that arise from the use of transitive dependencies can directly
influence the decision-making of APIs. For example, a user in PR \#20 of IDnow
SDK,\footnote{\ahref{http://web.archive.org/web/20201201224416/https://github.com/idnow/de.idnow.ios.sdk/issues/20}{https://github.com/idnow/de.idnow.ios.sdk/issues/20}}
an identity verification framework, is persuading the maintainers to drop
dependence on Sentry, an application monitoring platform, due to the user having
problems with Sentry as several versions of that dependency exist in its application. 

Given the increasing growth of indirect API calls and a slight increase of
multiple copies of the same function identifier appearing in clients, we call
for researchers to also account for the dynamics of dependency
management---particularly transitive dependencies---when studying the evolution
and decision-making around APIs. 
\end{sloppypar}

\section{Threats to Validity}\label{sec:threats}
In this section, we discuss limitations and threats that can affect the validity
of our study and show how we mitigated them.

\subsection{Internal validity}
\begin{sloppypar}
For CDNs to closely mirror actual package reuse in \crates, we only consider
packages specified under the \texttt{\#[dependencies]} section and
optionally-enabled packages as these are consumable in the source code. As
packages in \texttt{\#[dependencies]} are also available in the test portion of
packages, developers could potentially specify packages for testing purposes
that do not attribute towards package reuse. We mitigate the risk of inferring
test specific calls by restricting the build of packages to compilation without
further execution steps such as tests. 

The \texttt{rust-callgraphs} generator can resolve function invocations that
involve static and dynamic dispatch except for function pointer types. Although
the
documentation\footnote{\ahref{http://web.archive.org/web/20180416152826/https://doc.rust-lang.org/book/first-edition/trait-objects.html}{https://doc.rust-lang.org/book/first-edition/trait-objects.html}}
states that function pointers have a specific and limited purpose, we acknowledge
that we cannot make any claims around the completeness of generated CDNs due to
the general absence of ground truth for package repositories. When limiting the
scope to the features that the call graph generator supports, the generated CDNs
represent an over-approximation of function calls in \crates. It is an
over-approximation as function targets in dynamic dispatch may never be called
by the end-user in practice (i.e., it is inexact). Using additional analysis such
as dynamic analysis to remove all unlikely function targets is error-prone and
could result in unsound inferences. Thus, we avoid considering both static
(i.e., exact) and dynamic (i.e., inexact) function calls as the same. Instead,
we view the results of dynamically dispatched calls from the perspective of
virtual method tables (i.e., its concrete representation during runtime). 

Real-world constraints such as non-updated caches of the repository index,
user-defined dependency patches, and deviating semver specifications
could influence the actual version resolution of package dependencies. The
selection of packages and their versions for creating snapshots has additional
implications on the representativeness of \crates and its users. To mitigate the
risk of making incoherent versions resolutions, we use the exact resolver component implemented in \cargo, ensuring the
same treatment of version constraints. 

~\cite{kikas2017structure} report the highest package reach to be up
to 30\% in 2015 while our \crates metadata network report over 60\%, nearly
twice the number. The difference lies in the selection of packages when creating
the networks:~\cite{kikas2017structure} build a dependency tree for
all available versions of a package valid at timestamp $t$ and we build a tree for the
single most recent version of a package at a timestamp $t$. As there is no
consensus on best practices for which packages and releases to include in a
network, we take a conservative approach that avoids including dormant and
unused releases. For example, we argue that it is rare that a user today would
declare a dependence on version dating back to 2017 when newer versions from 2019
exist.~\cite{kikas2017structure} would include such versions.

\subsection{External and reliability validity}
We acknowledge that the results of network analysis are not generalizable to
other package repositories and only explain properties of \crates. Due to
differences in community values~\citep{bogart2016break} and reuse practices of
packages, we expect network analyses to yield different results. However, based
on~\citep{decan2018empirical} comparison of seven package repositories, we
believe certain repositories, for example, \npm and \nuget may share some
similarities with \crates than with \cran and \cpan.

The \ename approach to constructing a CDN is general applicability as long as
the programming language has a resolver for package dependencies and a call
graph generator. However, the soundness of generated CDNs may vary depending on
the programming language. For example, CDNs generated for Java are more accurate
and practical than CDNs for Python due to limited call graph support. Therefore,
evaluating trade-offs in terms of precision and recall plays an important role
in whether a study scenario is suitable for CDN analysis. 
\end{sloppypar}

\section{Future work}
Our work opens an array of opportunities for future work in data-driven analysis
of package repositories, both for researchers and tool builders. 

\subsection{Enabling data-driven insights into code reuse with network analysis}
As functions are not the only form of achieving code reuse, we aim to explore
how we can model reuse of interfaces, generics, class hierarchies, and wrapper
classes as networks. In a similar spirit to enabling data-driven insights of
APIs, language designers can use data-driven models to understand patterns and
adoption of certain code reuse practices. As Rust advocates developers to prefer
using generics over trait objects and limit the use of unsafe code constructs,
language designers can verify such premises with feedback through network- and
data-driven analyses of package repositories.

Following~\cite{zhang2020enabling}'s need-finding study on
data-driven API design, we are investigating possibilities to mine program
contexts and error-inducing patterns using \ename to extract API usage patterns
beyond syntactic features and frequencies. Insights into involved API usage
patterns can help library maintainers to make changes echoing improvements that
simplify code reuse and strengthening the stability of a package repository. 

\subsection{Modeling socio-technical risks of package abandonment}
Package repositories are successful in attracting developers to release new
packages. However, they are less successful in keeping these packages maintained
on a long-term perspective. As a result of developers abandoning packages due to
shifting priorities, unmaintained packages are increasingly jeopardizing the
security and stability of package repositories. Notably, the
\textit{event-stream} incident~\citep{npm:eventstream} is emerging as a textbook
example of how the abandonment of a package turned itself into a bitcoin stealing
apparatus affecting thousands of users. While survival analysis of packages can
yield insights into the stages of abandonment~\citep{valiev2018ecosystem},
understanding the social-technical motives behind developer abandonment could
potentially help develop a risk control model that package repository owners can
exercise. As an example, when a package repository recognizes the slowdown of
development activities of popular yet central packages, they could explore
incentives such as monetary support, developer assistant in resolving
long-running bug reports, or discuss possible handover to a network of trustful
developers. We are exploring both quantitative and qualitative strategies on how
to model and mitigate risks around package abandonment using \ename.

\section{Conclusions}
In this work, we devise \ename, an approach combining manifests and call graphs
of packages to infer dependency networks of package repositories at the function
granularity. By implementing \ename for Rust's \crates, we showcase the
feasibility of compiling and generating call graphs for 70\% of all indexed
releases. Then, we compare the \crates CDN against a conventional metadata-based
network and an enhanced corroborated version with compile data in a
study to understand their differences and similarities in network analysis
common to package repositories and derive new insights of \crates. By using
function call data, we find that packages do not indirectly call 60\% of their
transitive dependencies. Packages that have transitive consumers are
likely to have three times more calls from indirect users than direct users.
When we investigated the trends of function calls, we observed that 
packages make 6.6 new direct and 12.2 new indirect calls to
dependencies every six months.
A majority of packages in \crates have no or limited reachability; the
most reachable function in 2020---~\texttt{max-level()} in package
\texttt{log}---reaches 30\% of all \crates packages.
When comparing the three studied networks, we find that metadata-based networks
closely approximates the CDN for analysis involving direct package
relationships. Analysis of transitive package relationships and top-most
dependent packages, on the other hand, yield significantly different results for
the studied networks. A manual investigation of 34 cases reveals that a CDN is
more precise as it accounts for the context of how packages use
each other. Thus, dependency checkers such as, Rust's \texttt{cargo-audit} and
\github's \texttt{Dependabot}, can benefit from call graph analysis to generate
more precise recommendations for developers on transitive dependencies.
Overall, \ename opens up new doors to precise network analysis of code reuse and
APIs of package repositories.\\\\

\noindent
\textbf{Acknowledgments}: The work in this paper was partially funded by NWO grant 628.008.001 (CodeFeedr)
and H2020 grant 825328 (FASTEN).

\bibliographystyle{spbasic} 

\bibliography{praezi}   

\begin{thebibliography}{69}
\providecommand{\natexlab}[1]{#1}
\providecommand{\url}[1]{{#1}}
\providecommand{\urlprefix}{URL }
\expandafter\ifx\csname urlstyle\endcsname\relax
  \providecommand{\doi}[1]{DOI~\discretionary{}{}{}#1}\else
  \providecommand{\doi}{DOI~\discretionary{}{}{}\begingroup
  \urlstyle{rm}\Url}\fi
\providecommand{\eprint}[2][]{\url{#2}}

\bibitem[{Abdalkareem et~al.(2017)Abdalkareem, Nourry, Wehaibi, Mujahid, and
  Shihab}]{abdalkareem2017developers}
Abdalkareem R, Nourry O, Wehaibi S, Mujahid S, Shihab E (2017) Why do
  developers use trivial packages? an empirical case study on npm. In:
  Proceedings of the 2017 11th Joint Meeting on Foundations of Software
  Engineering, ACM, pp 385--395

\bibitem[{Abdalkareem et~al.(2019)Abdalkareem, Oda, Mujahid, and
  Shihab}]{abdalkareem2019impact}
Abdalkareem R, Oda V, Mujahid S, Shihab E (2019) On the impact of using trivial
  packages: an empirical case study on npm and pypi. Empirical Software
  Engineering pp 1--37

\bibitem[{Albert and Barab{\'a}si(2002)}]{albert2002statistical}
Albert R, Barab{\'a}si AL (2002) Statistical mechanics of complex networks.
  Reviews of modern physics 74(1):47

\bibitem[{Ali and Lhot{\'a}k(2012)}]{ali2012application}
Ali K, Lhot{\'a}k O (2012) Application-only call graph construction. In:
  European Conference on Object-Oriented Programming, Springer, pp 688--712

\bibitem[{Alimadadi et~al.(2015)Alimadadi, Mesbah, and
  Pattabiraman}]{alimadadi2015hybrid}
Alimadadi S, Mesbah A, Pattabiraman K (2015) Hybrid dom-sensitive change impact
  analysis for javascript. In: 29th European Conference on Object-Oriented
  Programming (ECOOP 2015), Schloss Dagstuhl-Leibniz-Zentrum fuer Informatik

\bibitem[{Aparicio(2019)}]{cargo:call:stack}
Aparicio J (2019) cargo-call-stack: Static, whole program stack analysis.
  \ahref{https://web.archive.org/web/20201202093844/https://github.com/japaric/cargo-call-stack}{https://github.com/japaric/cargo-call-stack}

\bibitem[{Backstrom et~al.(2012)Backstrom, Boldi, Rosa, Ugander, and
  Vigna}]{backstrom2012four}
Backstrom L, Boldi P, Rosa M, Ugander J, Vigna S (2012) Four degrees of
  separation. In: Proceedings of the 4th Annual ACM Web Science Conference, pp
  33--42

\bibitem[{Baldwin(2018)}]{npm:eventstream}
Baldwin A (2018) Details about the event-stream incident.
  \ahref{https://web.archive.org/web/20201202094027/https://blog.npmjs.org/post/180565383195/details-about-the-event-stream-incident}{https://blog.npmjs.org/post/180565383195/details-about-the-event-stream-incident}

\bibitem[{Beller et~al.(2016)Beller, Bholanath, McIntosh, and
  Zaidman}]{beller2016stateofasats}
Beller M, Bholanath R, McIntosh S, Zaidman A (2016) Analyzing the state of
  static analysis: A large-scale evaluation in open source software. In:
  Proceedings of the 23rd IEEE International Conference on Software Analysis,
  Evolution, and Reengineering, IEEE, pp 470--481

\bibitem[{Bogart et~al.(2016)Bogart, K{\"a}stner, Herbsleb, and
  Thung}]{bogart2016break}
Bogart C, K{\"a}stner C, Herbsleb J, Thung F (2016) How to break an {API}: Cost
  negotiation and community values in three software ecosystems. In:
  Proceedings of the 2016 24th ACM SIGSOFT International Symposium on
  Foundations of Software Engineering, ACM, pp 109--120

\bibitem[{Boldi et~al.(2004)Boldi, Codenotti, Santini, and
  Vigna}]{boldi2004ubicrawler}
Boldi P, Codenotti B, Santini M, Vigna S (2004) Ubicrawler: A scalable fully
  distributed web crawler. Software: Practice and Experience 34(8):711--726

\bibitem[{Brian et~al.(2020)Brian, David, and Aaron}]{rust:lib:blitz}
Brian A, David T, Aaron T (2020) The rust libz blitz.
  \ahref{https://web.archive.org/web/20201202094249/https://blog.rust-lang.org/2017/05/05/libz-blitz.html}{https://blog.rust-lang.org/2017/05/05/libz-blitz.html}

\bibitem[{Chen et~al.(2020)Chen, Hassan, Wang, and Zhang}]{chen2020taming}
Chen L, Hassan F, Wang X, Zhang L (2020) Taming behavioral backward
  incompatibilities via cross-project testing and analysis. In: IEEE/ACM
  International Conference on Software Engineering

\bibitem[{Chinthanet et~al.(2020)Chinthanet, Ponta, Plate, Sabetta, Kula,
  Ishio, and Matsumoto}]{chinthanet2020code}
Chinthanet B, Ponta SE, Plate H, Sabetta A, Kula RG, Ishio T, Matsumoto K
  (2020) Code-based vulnerability detection in node. js applications: How far
  are we? In: 2020 35th IEEE/ACM International Conference on Automated Software
  Engineering (ASE), IEEE, pp 1199--1203

\bibitem[{Cogo et~al.(2019)Cogo, Oliva, and Hassan}]{cogo2019empirical}
Cogo FR, Oliva GA, Hassan AE (2019) An empirical study of dependency downgrades
  in the npm ecosystem. IEEE Transactions on Software Engineering

\bibitem[{Decan et~al.(2018{\natexlab{a}})Decan, Mens, and
  Constantinou}]{decan2018impact}
Decan A, Mens T, Constantinou E (2018{\natexlab{a}}) On the impact of security
  vulnerabilities in the npm package dependency network. In: International
  Conference on Mining Software Repositories

\bibitem[{Decan et~al.(2018{\natexlab{b}})Decan, Mens, and
  Grosjean}]{decan2018empirical}
Decan A, Mens T, Grosjean P (2018{\natexlab{b}}) An empirical comparison of
  dependency network evolution in seven software packaging ecosystems.
  Empirical Software Engineering

\bibitem[{Decan et~al.(2019)Decan, Mens, and Grosjean}]{decan2019empirical}
Decan A, Mens T, Grosjean P (2019) An empirical comparison of dependency
  network evolution in seven software packaging ecosystems. Empirical Software
  Engineering 24(1):381--416

\bibitem[{Dietrich et~al.(2019)Dietrich, Pearce, Stringer, Tahir, and
  Blincoe}]{dietrich2019dependency}
Dietrich J, Pearce D, Stringer J, Tahir A, Blincoe K (2019) Dependency
  versioning in the wild. In: 2019 IEEE/ACM 16th International Conference on
  Mining Software Repositories (MSR), IEEE, pp 349--359

\bibitem[{Duan et~al.(2017)Duan, Bijlani, Xu, Kim, and
  Lee}]{duan2017identifying}
Duan R, Bijlani A, Xu M, Kim T, Lee W (2017) Identifying open-source license
  violation and 1-day security risk at large scale. In: Proceedings of the 2017
  ACM SIGSAC Conference on computer and communications security, pp 2169--2185

\bibitem[{Dunn(2017)}]{pypi:typo}
Dunn J (2017) Pypi python repository hit by typosquatting sneak attack.
  \ahref{https://web.archive.org/web/20201202094341/https://nakedsecurity.sophos.com/2017/09/19/pypi-python-repository-hit-by-typosquatting-sneak-attack/}{https://nakedsecurity.sophos.com/2017/09/19/pypi-python-repository-hit-by-typosquatting-sneak-attack/}

\bibitem[{Emami et~al.(1994)Emami, Ghiya, and Hendren}]{emami1994context}
Emami M, Ghiya R, Hendren LJ (1994) Context-sensitive interprocedural points-to
  analysis in the presence of function pointers. ACM SIGPLAN Notices
  29(6):242--256

\bibitem[{Hejderup(2015)}]{hejderup2015dependencies}
Hejderup J (2015) In dependencies we trust: How vulnerable are dependencies in
  software modules? Master's thesis, Delft University of technology

\bibitem[{Hejderup et~al.(2018)Hejderup, van Deursen, and
  Gousios}]{hejderup2018software}
Hejderup J, van Deursen A, Gousios G (2018) Software ecosystem call graph for
  dependency management. In: Proceedings of the 40th International Conference
  on Software Engineering: New Ideas and Emerging Results, ACM, pp 101--104

\bibitem[{Hejderup et~al.(2021)Hejderup, Beller, Triantafyllou, and
  Gousios}]{joseph_hejderup_2021_4478981}
Hejderup J, Beller M, Triantafyllou K, Gousios G (2021) {Präzi: From
  Package-based to Call-based Dependency Networks}.
  \doi{10.5281/zenodo.4478981},
  \urlprefix\url{https://doi.org/10.5281/zenodo.4478981}

\bibitem[{Hopkins(1997)}]{hopkins1997new}
Hopkins WG (1997) A new view of statistics. Will G. Hopkins

\bibitem[{Katz(2016)}]{blogrustcargo}
Katz Y (2016) Cargo: predictable dependency management.
  \ahref{https://web.archive.org/web/20180508173027/https://blog.rust-lang.org/2016/05/05/cargo-pillars.html}{https://blog.rust-lang.org/2016/05/05/cargo-pillars.html}

\bibitem[{Kikas et~al.(2017)Kikas, Gousios, Dumas, and
  Pfahl}]{kikas2017structure}
Kikas R, Gousios G, Dumas M, Pfahl D (2017) Structure and evolution of package
  dependency networks. In: Proceedings of the 14th International Conference on
  Mining Software Repositories, IEEE Press, pp 102--112

\bibitem[{Kula et~al.(2018{\natexlab{a}})Kula, De~Roover, German, Ishio, and
  Inoue}]{kula2017modeling}
Kula RG, De~Roover C, German DM, Ishio T, Inoue K (2018{\natexlab{a}}) A
  generalized model for visualizing library popularity, adoption, and diffusion
  within a software ecosystem. In: 2018 IEEE 25th International Conference on
  Software Analysis, Evolution and Reengineering (SANER), IEEE, pp 288--299

\bibitem[{Kula et~al.(2018{\natexlab{b}})Kula, Ouni, German, and
  Inoue}]{kula2018empirical}
Kula RG, Ouni A, German DM, Inoue K (2018{\natexlab{b}}) An empirical study on
  the impact of refactoring activities on evolving client-used apis.
  Information and Software Technology 93:186--199

\bibitem[{Lehman(1980)}]{lehman1980programs}
Lehman MM (1980) Programs, life cycles, and laws of software evolution.
  Proceedings of the IEEE 68(9):1060--1076

\bibitem[{Livshits et~al.(2015)Livshits, Sridharan, Smaragdakis, Lhot{\'a}k,
  Amaral, Chang, Guyer, Khedker, M{\o}ller, and
  Vardoulakis}]{livshits2015defense}
Livshits B, Sridharan M, Smaragdakis Y, Lhot{\'a}k O, Amaral JN, Chang BYE,
  Guyer SZ, Khedker UP, M{\o}ller A, Vardoulakis D (2015) In defense of
  soundiness: a manifesto. Communications of the ACM 58(2):44--46

\bibitem[{Martins et~al.(2018)Martins, Achar, and Lopes}]{martins201850k}
Martins P, Achar R, Lopes CV (2018) 50k-c: A dataset of compilable, and
  compiled, java projects. In: 2018 IEEE/ACM 15th International Conference on
  Mining Software Repositories (MSR), IEEE, pp 1--5

\bibitem[{Matsakis(2016)}]{rustblog:mir}
Matsakis N (2016) Introducing mir.
  \ahref{https://web.archive.org/web/20201202094745/https://blog.rust-lang.org/2016/04/19/MIR.html}{https://blog.rust-lang.org/2016/04/19/MIR.html}

\bibitem[{Mezzetti et~al.(2018)Mezzetti, M{\o}ller, and
  Torp}]{mezzetti2018type}
Mezzetti G, M{\o}ller A, Torp MT (2018) Type regression testing to detect
  breaking changes in node. js libraries. In: 32nd European Conference on
  Object-Oriented Programming (ECOOP 2018), Schloss Dagstuhl-Leibniz-Zentrum
  fuer Informatik

\bibitem[{Mirhosseini and Parnin(2017)}]{mirhosseini2017can}
Mirhosseini S, Parnin C (2017) Can automated pull requests encourage software
  developers to upgrade out-of-date dependencies? In: Proceedings of the 32nd
  IEEE/ACM International Conference on Automated Software Engineering, IEEE
  Press, pp 84--94

\bibitem[{Mones et~al.(2012)Mones, Vicsek, and Vicsek}]{mones2012hierarchy}
Mones E, Vicsek L, Vicsek T (2012) Hierarchy measure for complex networks. PloS
  one 7(3):e33799

\bibitem[{Mujahid et~al.(2020)Mujahid, Abdalkareem, Shihab, and
  McIntosh}]{Mujahid_MSR2020}
Mujahid S, Abdalkareem R, Shihab E, McIntosh S (2020) Using others' tests to
  identify breaking updates. In: Proceedings of the 17th International
  Conference on Mining Software Repositories, pp 466--476

\bibitem[{Nguyen et~al.(2019)Nguyen, Nguyen, Dig, Nguyen, Tran, and
  Hilton}]{nguyen2019graph}
Nguyen HA, Nguyen TN, Dig D, Nguyen S, Tran H, Hilton M (2019) Graph-based
  mining of in-the-wild, fine-grained, semantic code change patterns. In: 2019
  IEEE/ACM 41st International Conference on Software Engineering (ICSE), IEEE,
  pp 819--830

\bibitem[{Ponta et~al.(2018)Ponta, Plate, and Sabetta}]{ponta2018beyond}
Ponta SE, Plate H, Sabetta A (2018) Beyond metadata: Code-centric and
  usage-based analysis of known vulnerabilities in open-source software. In:
  2018 IEEE International Conference on Software Maintenance and Evolution
  (ICSME), IEEE, pp 449--460

\bibitem[{Porter et~al.(2009)Porter, Onnela, and Mucha}]{porter2009communities}
Porter MA, Onnela JP, Mucha PJ (2009) Communities in networks. Notices of the
  AMS 56(9):1082--1097

\bibitem[{Preston-Werner(2013)}]{semver}
Preston-Werner T (2013) Semantic versioning.
  \ahref{https://web.archive.org/web/20201201220024/https://semver.org/}{https://semver.org/}

\bibitem[{Raemaekers et~al.(2017)Raemaekers, van Deursen, and
  Visser}]{raemaekers2017semantic}
Raemaekers S, van Deursen A, Visser J (2017) Semantic versioning and impact of
  breaking changes in the maven repository. Journal of Systems and Software
  129:140--158

\bibitem[{Robbes et~al.(2012)Robbes, Lungu, and
  R{\"o}thlisberger}]{robbes2012developers}
Robbes R, Lungu M, R{\"o}thlisberger D (2012) How do developers react to api
  deprecation? the case of a smalltalk ecosystem. In: Proceedings of the ACM
  SIGSOFT 20th International Symposium on the Foundations of Software
  Engineering, pp 1--11

\bibitem[{Ryder(1979)}]{ryder1979constructing}
Ryder BG (1979) Constructing the call graph of a program. IEEE Transactions on
  Software Engineering (3):216--226

\bibitem[{Salis et~al.(2021)Salis, Sotiropoulos, Louridas, Spinellis, and
  Mitropoulos}]{salis2021pycg}
Salis V, Sotiropoulos T, Louridas P, Spinellis D, Mitropoulos D (2021) Pycg:
  Practical call graph generation in python. In: 2021 IEEE/ACM 43rd
  International Conference on Software Engineering (ICSE), IEEE, pp 1646--1657

\bibitem[{Sawant and Bacchelli(2017)}]{sawant2017fine}
Sawant AA, Bacchelli A (2017) fine-grape: fine-grained api usage extractor--an
  approach and dataset to investigate api usage. Empirical Software Engineering
  22(3):1348--1371

\bibitem[{Sawant et~al.(2018{\natexlab{a}})Sawant, Aniche, van Deursen, and
  Bacchelli}]{sawant2018understanding}
Sawant AA, Aniche M, van Deursen A, Bacchelli A (2018{\natexlab{a}})
  Understanding developers' needs on deprecation as a language feature. In:
  2018 IEEE/ACM 40th International Conference on Software Engineering (ICSE),
  IEEE, pp 561--571

\bibitem[{Sawant et~al.(2018{\natexlab{b}})Sawant, Aniche, van Deursen, and
  Bacchelli}]{Sawantdeprecation}
Sawant AA, Aniche M, van Deursen A, Bacchelli A (2018{\natexlab{b}})
  Understanding developers' needs on deprecation as a language feature. In:
  Proceedings of the 40th International Conference on Software Engineering,
  ACM, New York, NY, USA, ICSE '18, pp 561--571

\bibitem[{Sawant et~al.(2018{\natexlab{c}})Sawant, Robbes, and
  Bacchelli}]{sawant2018reaction}
Sawant AA, Robbes R, Bacchelli A (2018{\natexlab{c}}) On the reaction to
  deprecation of clients of 4+ 1 popular java apis and the jdk. Empirical
  Software Engineering 23(4):2158--2197

\bibitem[{Schlueter(2013)}]{npm:unix:design}
Schlueter I (2013) Unix philosophy and node.js.
  \ahref{https://web.archive.org/web/20200503155523/https://blog.izs.me/2013/04/unix-philosophy-and-nodejs/}{https://blog.izs.me/2013/04/unix-philosophy-and-nodejs}

\bibitem[{Schlueter(2017)}]{npm:leftpad:online}
Schlueter I (2017) The npm blog — kik, left-pad, and npm.
  \ahref{http://web.archive.org/web/20180416152826/http://blog.npmjs.org/post/141577284765/kik-left-pad-and-npm}{http://blog.npmjs.org/post/141577284765/kik-left-pad-and-npm}

\bibitem[{Shapiro and Horwitz(1997)}]{shapiro1997fast}
Shapiro M, Horwitz S (1997) Fast and accurate flow-insensitive points-to
  analysis. In: Proceedings of the 24th ACM SIGPLAN-SIGACT symposium on
  Principles of programming languages, pp 1--14

\bibitem[{Shivers(1991)}]{shivers1991control}
Shivers O (1991) Control-flow analysis of higher-order languages. PhD thesis,
  PhD thesis, Carnegie Mellon University

\bibitem[{Steensgaard(1996)}]{steensgaard1996points}
Steensgaard B (1996) Points-to analysis in almost linear time. In: Proceedings
  of the 23rd ACM SIGPLAN-SIGACT symposium on Principles of programming
  languages, pp 32--41

\bibitem[{Sul{\'\i}r and Porub{\"a}n(2016)}]{sulir2016quantitative}
Sul{\'\i}r M, Porub{\"a}n J (2016) A quantitative study of {Java} software
  buildability. In: Proceedings of the 7th International Workshop on Evaluation
  and Usability of Programming Languages and Tools, ACM, pp 17--25

\bibitem[{Sundaresan et~al.(2000)Sundaresan, Hendren, Razafimahefa,
  Vall{\'e}e-Rai, Lam, Gagnon, and Godin}]{sundaresan2000practical}
Sundaresan V, Hendren L, Razafimahefa C, Vall{\'e}e-Rai R, Lam P, Gagnon E,
  Godin C (2000) Practical virtual method call resolution for Java, vol~35. ACM

\bibitem[{Tip and Palsberg(2000)}]{tip2000scalable}
Tip F, Palsberg J (2000) Scalable propagation-based call graph construction
  algorithms. In: Proceedings of the 15th ACM SIGPLAN conference on
  Object-oriented programming, systems, languages, and applications, pp
  281--293

\bibitem[{Triantafyllou(2019)}]{rustforum:cgbench}
Triantafyllou K (2019) A benchmark for rust call-graph generators.
  \ahref{https://web.archive.org/web/20201202095033/https://users.rust-lang.org/t/a-benchmark-for-rust-call-graph-generators/34494}{https://users.rust-lang.org/t/a-benchmark-for-rust-call-graph-generators/34494}

\bibitem[{Tufano et~al.(2017)Tufano, Palomba, Bavota, Di~Penta, Oliveto,
  De~Lucia, and Poshyvanyk}]{tufano2017there}
Tufano M, Palomba F, Bavota G, Di~Penta M, Oliveto R, De~Lucia A, Poshyvanyk D
  (2017) There and back again: Can you compile that snapshot? Journal of
  Software: Evolution and Process 29(4)

\bibitem[{Valiev et~al.(2018)Valiev, Vasilescu, and
  Herbsleb}]{valiev2018ecosystem}
Valiev M, Vasilescu B, Herbsleb J (2018) Ecosystem-level determinants of
  sustained activity in open-source projects: a case study of the pypi
  ecosystem. In: Proceedings of the 2018 26th ACM Joint Meeting on European
  Software Engineering Conference and Symposium on the Foundations of Software
  Engineering, ACM, pp 644--655

\bibitem[{Wittern et~al.(2016)Wittern, Suter, and
  Rajagopalan}]{wittern2016look}
Wittern E, Suter P, Rajagopalan S (2016) A look at the dynamics of the
  {J}ava{S}cript package ecosystem. In: Mining Software Repositories (MSR),
  2016 IEEE/ACM 13th Working Conference on, IEEE, pp 351--361

\bibitem[{Xavier et~al.(2017)Xavier, Brito, Hora, and
  Valente}]{xavier2017historical}
Xavier L, Brito A, Hora A, Valente MT (2017) Historical and impact analysis of
  api breaking changes: A large-scale study. In: 2017 IEEE 24th International
  Conference on Software Analysis, Evolution and Reengineering (SANER), IEEE,
  pp 138--147

\bibitem[{Zapata et~al.(2018)Zapata, Kula, Chinthanet, Ishio, Matsumoto, and
  Ihara}]{zapata2018towards}
Zapata RE, Kula RG, Chinthanet B, Ishio T, Matsumoto K, Ihara A (2018) Towards
  smoother library migrations: A look at vulnerable dependency migrations at
  function level for npm javascript packages. In: 2018 IEEE International
  Conference on Software Maintenance and Evolution (ICSME), IEEE, pp 559--563

\bibitem[{Zerouali et~al.(2018)Zerouali, Constantinou, Mens, Robles, and
  Gonz{\'a}lez-Barahona}]{zerouali2018empirical}
Zerouali A, Constantinou E, Mens T, Robles G, Gonz{\'a}lez-Barahona J (2018) An
  empirical analysis of technical lag in npm package dependencies. In:
  International Conference on Software Reuse, Springer, pp 95--110

\bibitem[{Zhang et~al.(2020{\natexlab{a}})Zhang, Hartmann, Kim, and
  Glassman}]{zhang2020enabling}
Zhang T, Hartmann B, Kim M, Glassman EL (2020{\natexlab{a}}) Enabling
  data-driven api design with community usage data: A need-finding study. In:
  Proceedings of the 2020 CHI Conference on Human Factors in Computing Systems,
  pp 1--13

\bibitem[{Zhang et~al.(2020{\natexlab{b}})Zhang, Hartmann, Kim, and
  Glassman}]{zhangenabling}
Zhang T, Hartmann B, Kim M, Glassman EL (2020{\natexlab{b}}) Enabling
  data-driven api design with community usage data: A need-finding study. In:
  Proceedings of the 2020 CHI Conference on Human Factors in Computing Systems,
  pp 1--13

\bibitem[{Zhong et~al.(2010)Zhong, Thummalapenta, Xie, Zhang, and
  Wang}]{zhong2010mining}
Zhong H, Thummalapenta S, Xie T, Zhang L, Wang Q (2010) Mining api mapping for
  language migration. In: Proceedings of the 32nd ACM/IEEE International
  Conference on Software Engineering-Volume 1, pp 195--204

\bibitem[{Zimmermann et~al.(2019)Zimmermann, Staicu, Tenny, and
  Pradel}]{zimmermann2019small}
Zimmermann M, Staicu CA, Tenny C, Pradel M (2019) Small world with high risks:
  A study of security threats in the npm ecosystem. In: 28th USENIX Security
  Symposium (USENIX Security 19), pp 995--1010

\end{thebibliography}

\appendix

\section{Selecting a time window for dependency resolution}\label{apdix:verres}

\begin{figure*}[b]
\centering
\includegraphics[width=0.6\columnwidth,keepaspectratio]{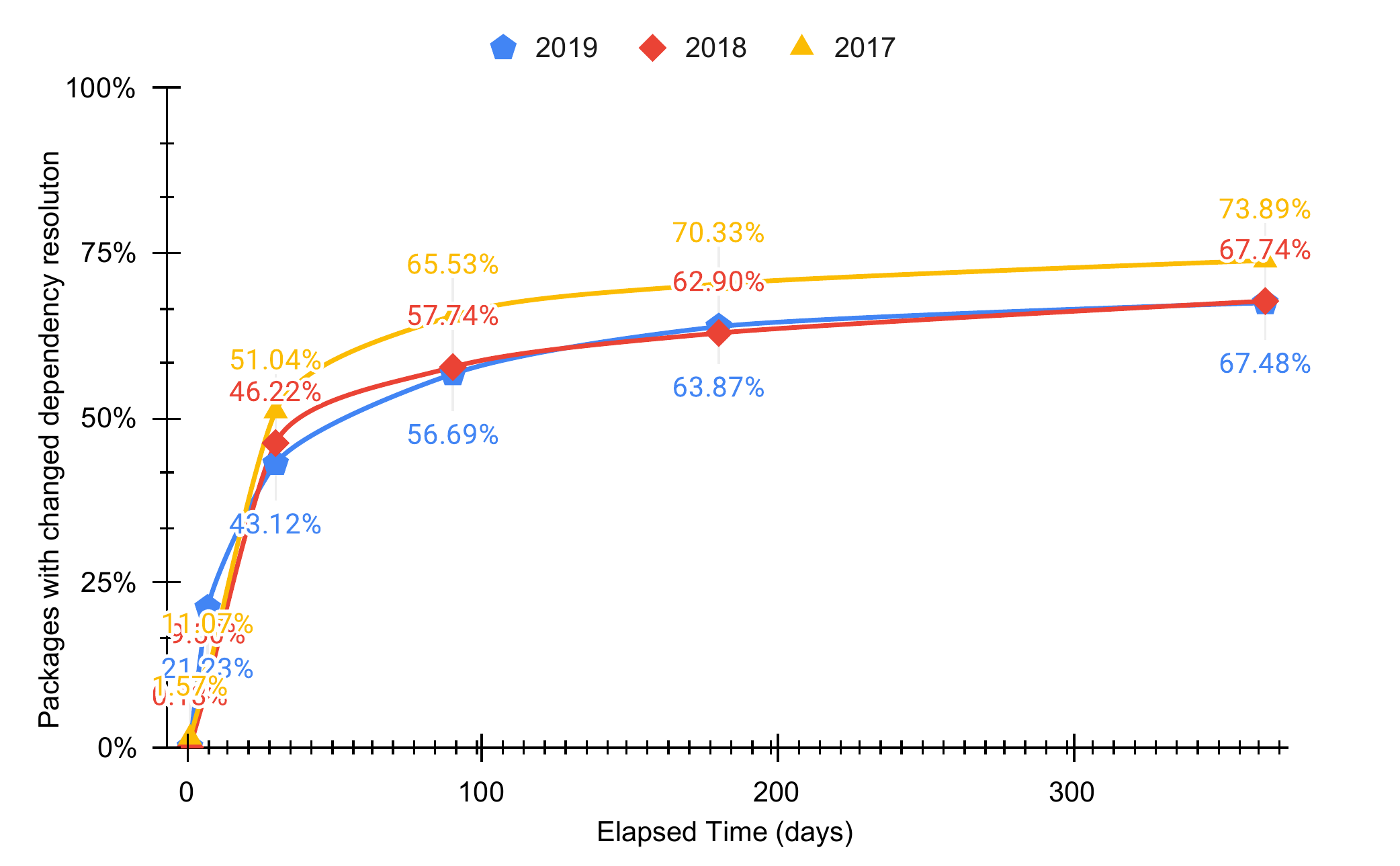}
\caption{Retroactive resolution of dependencies over a time period of one year in 2017, 2018, and 2019}
\label{fig:pdn_compare}
\end{figure*}

Instead of using a single fixed version at all times, version constraints allow
developers to use a time-constrained version that updates itself at new
compilations. Nearly all dependencies in \crates specify a dynamic version
constraint---only 2.92\% of all dependency specifications in \crates use a
single (immutable) version~\citep{dietrich2019dependency}. Before studying the evolution and structure of \crates, we first decide the number of time points and a time
window between each time point. Although popular studies such as~\citep{kikas2017structure}
and~\citep{decan2019empirical} use a time window of one year to study structural changes, we,
instead, determine a time window based on the frequency of structural changes in
\crates.

After resolving the dependency tree of a set of packages in \crates at a time
$t$, we then re-resolve it using six different time points (i.e., one day, one
week, one month, three months, six months, and one year) to find a time window
where a large fraction of them have a changed dependency tree. We perform this
using a set of packages having at least one non-optional dependency at the
beginning of 2017 (5,252 package releases), 2018 (9,716 package releases), and
2019 (16,098 package releases).

\Cref{fig:pdn_compare} shows the fraction of packages with a changed dependency
tree (i.e., a tree with at least one different version) over time. We observe a logarithmic trendline for each year group; a high
increase of packages with changed dependency between time points before three
months, and then it levels out. After one month, we already find that 40\% of
all packages have a changed dependency tree due to new releases of 148 packages
in 2017, 190 packages in 2018, and 240 packages in 2019. In all year groups, we
find that the dependence on \texttt{libc} triggers a new version resolution for most
packages, followed by other popular packages such as \texttt{quote},
\texttt{serde}, and \texttt{syn}. A manual inspection of the release log for
\texttt{libc}\footnote{\url{https://crates.io/crates/libc/versions}} and
\texttt{serde}\footnote{\url{https://crates.io/crates/serde/versions}}, suggests
a frequency of at least two releases per month.

Finally, we also observe that 26\% of all packages in 2017 have an identical
dependency tree after one year. Among those unchanged packages, nearly all of
them (2017: 83\%, 2018: 93\%, 2019: 90\%) are outdated packages. With outdated,
we mean that no recent releases for those packages in more than one
year. Although
packages may be outdated, they still could use flexible version constraints. In
roughly one-third (2017: 31\%, 2018: 34\% 2019: 40\%) of all dependency
constraints, the dependencies are outdated packages (i.e., there are no recent
releases). In the remaining cases (i.e., where more recent versions exist), the
version constraints cover old releases  (e.g., depending on \texttt{serde}
2.x when 4.x exists), and less than 1\% are fixed versions. For example,
\texttt{xml-attributes-derive::0.1.0}\footnote{\url{https://docs.rs/crate/xml-attributes-derive/0.1.0/source/Cargo.toml}}
depends on older
versions of \texttt{syn}, \texttt{quote}, and \texttt{proc-macro2}, and
\texttt{trie-root::0.11.0}\footnote{\url{https://docs.rs/crate/trie-root/0.11.0/source/Cargo.toml}} depends on an old version of \texttt{hash-db}.

Given these observations, we select a time window of one month and thus perform dependency resolution every month per year.

\begin{mdframed}[roundcorner=0pt,nobreak=true,align=center]
40\% of all \crates packages have at least one dependency resolving to a new version after 30 days.
\end{mdframed}

\end{document}